\pdfoutput=1

\documentclass[11pt]{article}

\usepackage{acl}

\usepackage{times}
\usepackage{latexsym}
\usepackage[T1]{fontenc}

\usepackage[utf8]{inputenc}

\usepackage{microtype}
\usepackage{float} 

\usepackage{inconsolata}

\usepackage{graphicx}
\usepackage{amssymb}
\usepackage{amsmath}
\usepackage{booktabs}
\usepackage{subcaption}
\usepackage{multirow}

\expandafter\def\expandafter\normalsize\expandafter{%
    \normalsize%
    \setlength\abovedisplayskip{4pt}%
    \setlength\belowdisplayskip{4pt}%
    \setlength\abovedisplayshortskip{0pt}%
    \setlength\belowdisplayshortskip{2pt}%
}

\title{\includegraphics[height=15pt]{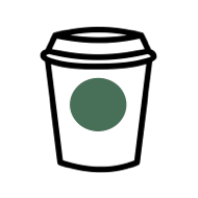} Starbucks: Improved Training for 2D Matryoshka Embeddings}

\author{
	Shengyao Zhuang\textsuperscript{1}\thanks{\textsuperscript{*}These authors contributed equally to this work.},
	Shuai Wang\textsuperscript{2}\footnotemark[1],
	Fabio Zheng\textsuperscript{2},
	Bevan Koopman\textsuperscript{1,2},
	Guido Zuccon\textsuperscript{2} \\
	\textsuperscript{1}CSIRO,\ \textsuperscript{2}The University of Queensland,\ Australia \\
	\texttt{\{shengyao.zhuang, b.koopman\}@csiro.au},\\
	\texttt{fabio.zheng@student.uq.edu.au},\\
	\texttt{\{shuai.wang2, g.zuccon\}@uq.edu.au}
}

\begin{document}
\maketitle
\begin{abstract}

2D Matryoshka training enables a single embedding model to generate sub-network representations across different layers and embedding dimensions, offering adaptability to diverse computational and task constraints. However, its effectiveness remains well below that of individually trained models of equivalent sizes.
To address this, we propose \textbf{Starbucks}, a new training strategy for Matryoshka-style embedding models that combines structured fine-tuning with masked autoencoder (MAE) pre-training. During fine-tuning, we compute the loss over a fixed set of layer-dimension pairs, from small to large, which significantly improves performance over randomly sampled sub-networks and matches that of separately trained models. Our MAE-based pre-training further enhances the representation quality of sub-networks, providing a stronger backbone for downstream tasks.
Experiments on both in-domain (semantic similarity and passage retrieval) and out-of-domain (BEIR) benchmarks show that Starbucks consistently outperforms 2D Matryoshka models and matches or exceeds the performance of individually trained models, while maintaining high efficiency and adaptability. Ablation studies confirm our loss design choices, the impact of SMAE pre-training and demonstrate the applicability of Starbucks across backbones. We further show that depth- and width-wise Starbucks variants capture complementary information, and that their hybridization yields additional performance gains with minimal latency overhead due to parallelization.%
\footnote{Code at \href{https://github.com/ielab/Starbucks}{github.com/ielab/Starbucks}.}

\end{abstract}

\section{Introduction}
\begin{figure}
	\centering
	\includegraphics[width=\columnwidth]{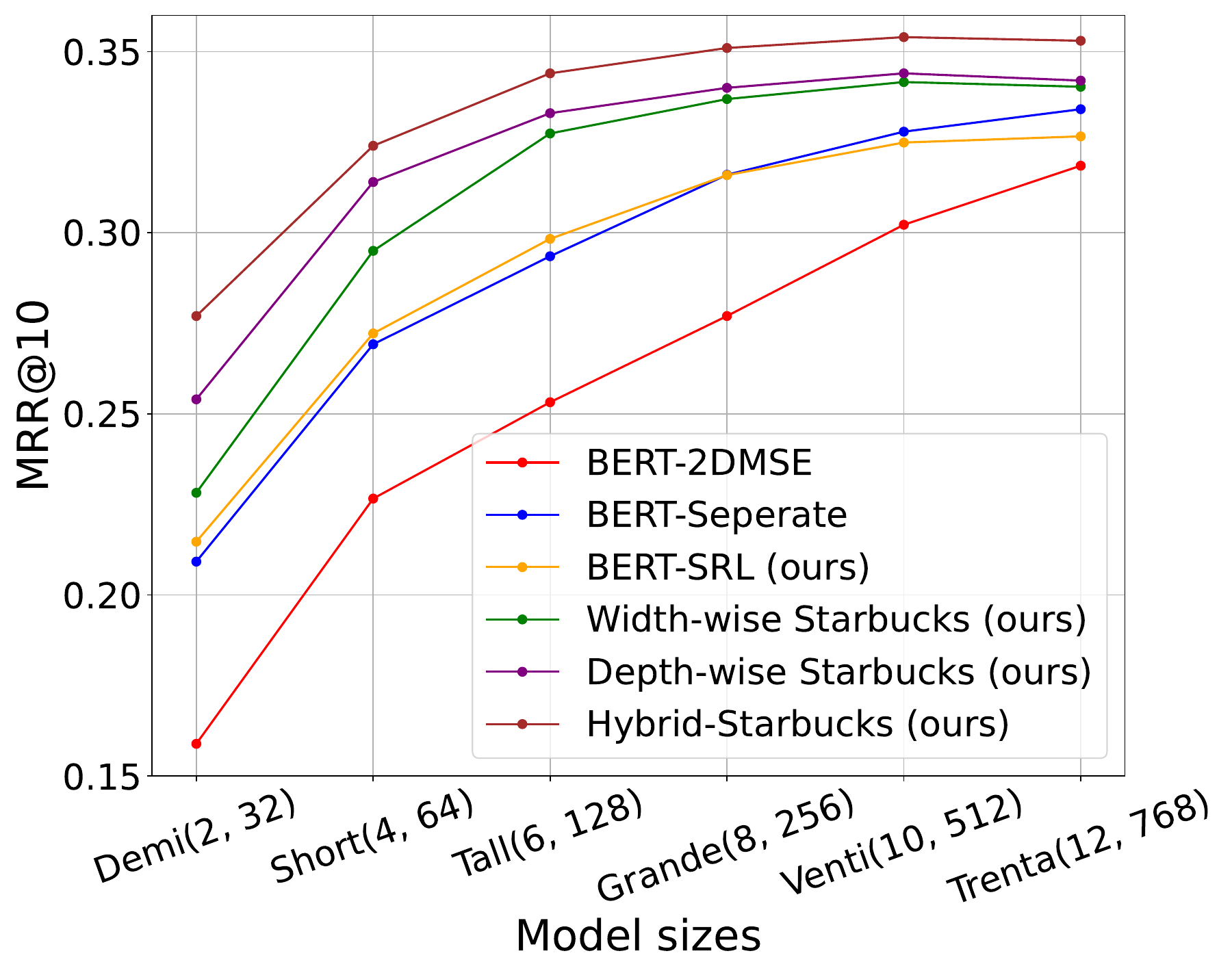}
	\caption{
		MRR@10 on MARCO dev obtained by different 2D Matryoshka training methods. The x-axis refers to model sizes in the format \textit{SizeName (number of layers, embedding dimension)} for Width-wise Starbucks. Depth-wise models use approximately the same number of parameters (difference <0.05\%) as their corresponding sizes.
	}
	\label{fig:introplot}
\end{figure}

\begin{figure*}[t!]
	\centering
	\begin{subfigure}[b]{0.2903\linewidth}
		\includegraphics[width=\linewidth]{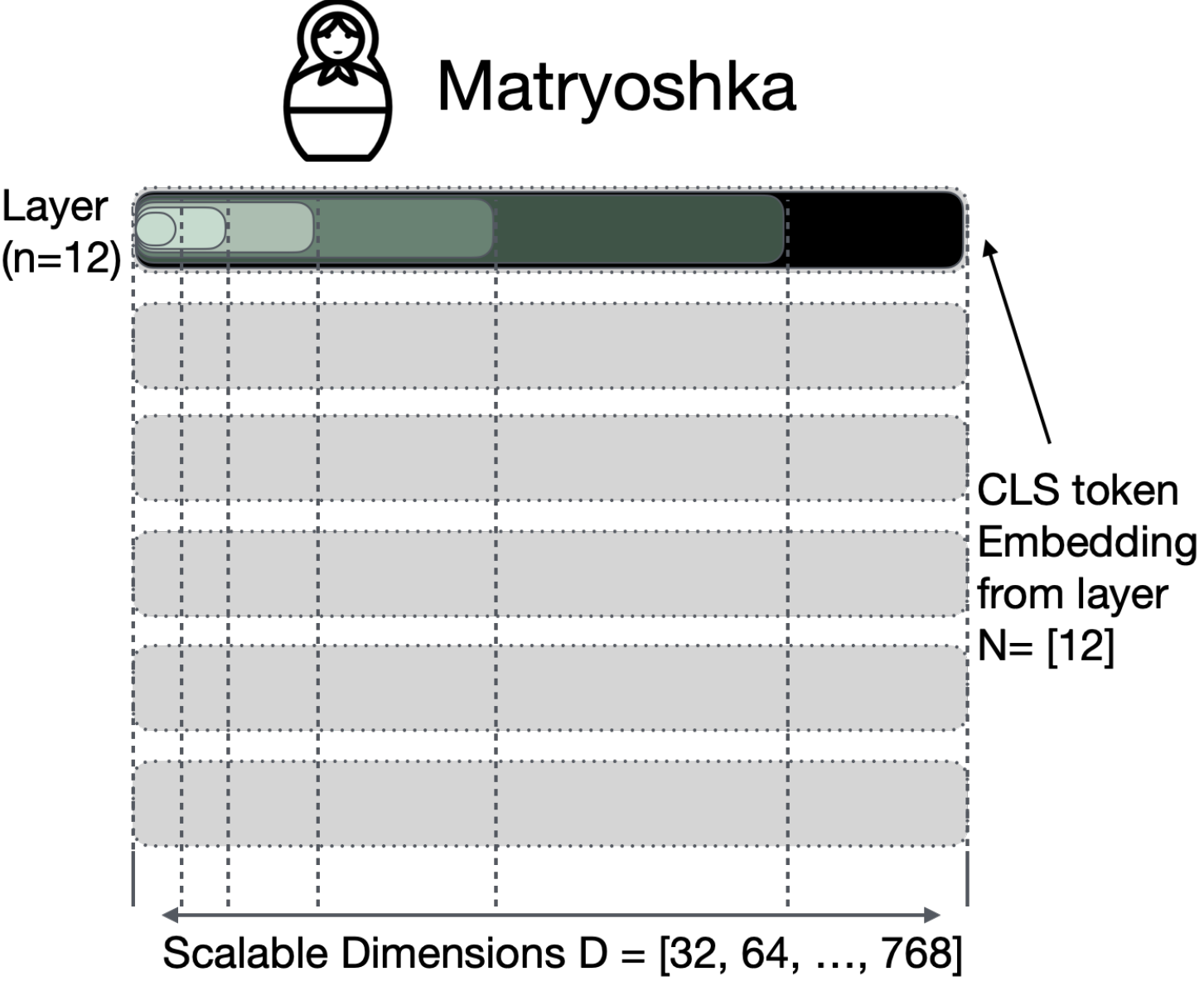}
		\label{fig:1dm}
	\end{subfigure}
	\hfill
	\begin{subfigure}[b]{0.3168\linewidth}
		\includegraphics[width=\linewidth]{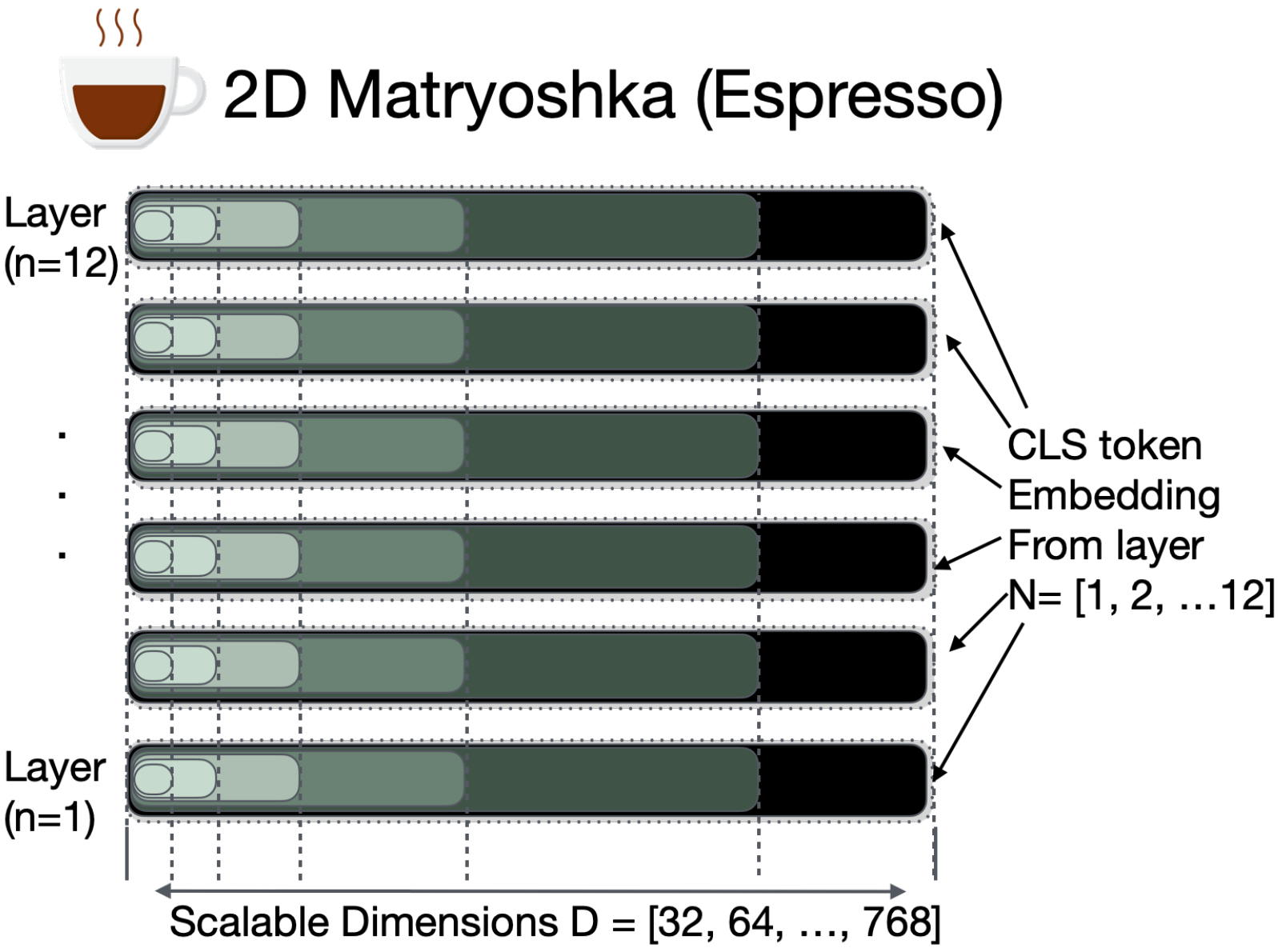}
		\label{fig:2dm}
	\end{subfigure}
	\hfill
	\begin{subfigure}[b]{0.373\linewidth}
		\includegraphics[width=\linewidth]{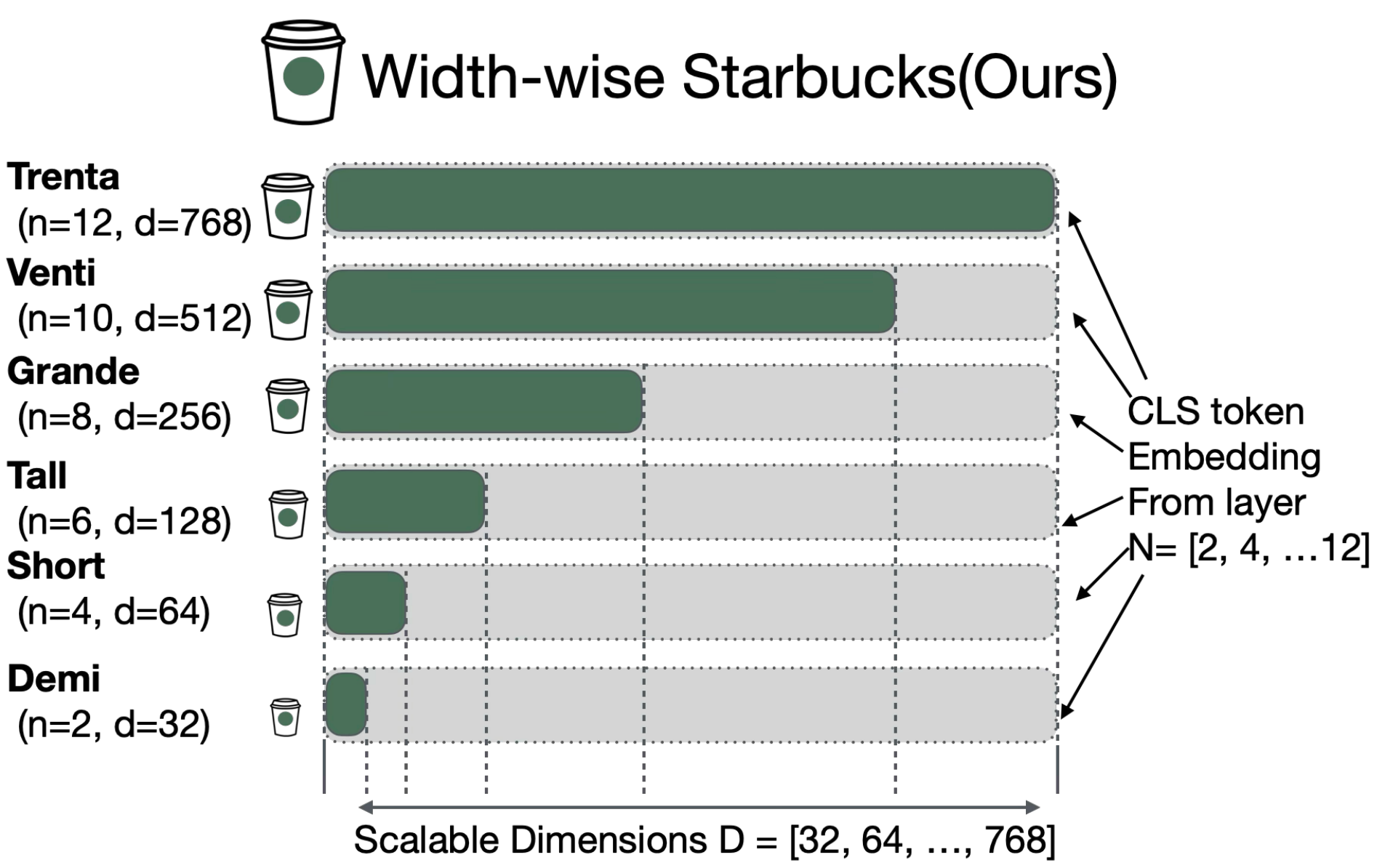}
		\label{fig:starbucks}
	\end{subfigure}
	\caption{
		Starbucks (right) provides a single model with flexible configurations in terms of embedding dimensionality and network depth. Also shown are the original Matryoshka model (left) and its extension, 2D Matryoshka (center). Here, we only use Width-wise Starbucks to illustrate the approach.
	}
	\label{fig:archi}
\end{figure*}

Matryoshka representation learning (MRL)~\cite{mrl2022} is a training technique that enables embedding models to support multiple dimensional granularities, allowing adaptability to different downstream resource constraints. Recently, the 2D Matryoshka Sentence Embeddings (2DMSE)\footnote{Also known as Espresso Sentence Embeddings (ESE).} method was proposed to further enhance flexibility by enabling elastic embedding generation across both embedding dimensions and model layers~\cite{li2024eseespressosentenceembeddings}. This approach allows a single embedding model to replace multiple models trained with different size configurations, reducing training and deployment costs.

However, prior work has shown that the effectiveness of embeddings generated by 2DMSE sub-networks (using fewer layers or lower dimensions) is significantly lower to that of individually trained models targeting the same sizes~\cite{wang20242d}. Figure~\ref{fig:introplot} confirms this degradation, highlighting a key limitation in existing 2DMSE training.
 In contrast, separately training multiple small-scale sub-models, despite higher training cost and inference memory use, remains preferred due to (1) significantly higher embedding quality, (2) the one-time nature of training, and (3) the limited number of configurations typically required in practice.

To address this, we propose \textbf{Starbucks}, a new 2DMSE training framework comprising two components: \textbf{Starbucks Masked Autoencoding} (SMAE) for pre-training and \textbf{Starbucks Representation Learning} (SRL) for fine-tuning. Starbucks computes loss over a curated set of target layer-dimension pairs, covering a spectrum of subnetwork sizes.
\textbf{SRL} promotes effective learning across these configurations—named after Starbucks cup sizes: Demi, Short, Tall, Grande, Venti, and Trenta (see Figure~\ref{fig:archi}). Our results show that Starbucks matches the performance of separately trained models, making Matryoshka learning practical.
\textbf{SMAE} integrates ideas from MatFormer~\cite{devvrit2023matformernestedtransformerelastic} and MAE-style embedding pre-training~\cite{gao-callan-2021-condenser,xiao-etal-2022-retromae,wang-etal-2023-simlm,zhuang2023typo,lu2021less}, applying masked autoencoding to representations from multiple layer-dimension pairs to prepare the model for SRL fine-tuning.

We evaluate Starbucks against 2DMSE and separately trained models on both in-domain (STS and passage retrieval) and out-of-domain (BEIR) tasks. Results show that: (1) Starbucks significantly improves over 2DMSE and even outperform the separately trained models while being more training time and inference memory efficient; (2) it generalizes well across architectures (BERT, E5); and (3) the training strategy supports both depth-wise and width-wise configurations. These variants produce complementary embeddings, especially in smaller subnetworks, and their hybridization yields further gains. As the configurations are independent, they can be applied in parallel, enabling scalable and efficient elastic embedding generation.

\section{Related Work}
Transformer-based text embedding models are highly effective across various NLP tasks, such as semantic text similarity~\cite{muennighoff-etal-2023-mteb,reimers-gurevych-2019-sentence,gao-etal-2021-simcse,li-li-2024-aoe} and passage retrieval~\cite{beir,karpukhin-etal-2020-dense,DRbuidu,zhuang2024promptrepspromptinglargelanguage}
The MRL approach aims to make embedding models more efficient and practical by utilising  sub-dimensions of the embeddings to calculate the training loss~\cite{mrl2022}. 
 After MRL training, all corresponding embeddings sub-dimensions can effectively represent the text, enabling more efficient inference for applications~\cite{lee2024geckoversatiletextembeddings} with adaptive algorithms~\cite{adanns}.

The original MRL method enabled adaptability only across output embedding dimensions. 2DMSE extends this by incorporating embeddings from multiple model layers, allowing adaptation to both embedding dimensions and model sizes (i.e., number of layers)~\cite{li2024eseespressosentenceembeddings}. 
Our proposed \textbf{Starbucks} similarly supports elastic embedding across dimensions and layers, but focuses on a fixed set of layer-dimension pairs from small to large. We show that this targeted training strategy is more effective than 2DMSE. Figure~\ref{fig:archi} illustrates the architectural differences between Starbucks and prior MRL approaches.

In addition to MRL for embedding models, other works have explored nested or elastic architectures in unsupervised language modeling. For instance, MatFormer introduced a nested transformer block for decoder-only language models~\cite{devvrit2023matformernestedtransformerelastic}. Drawing inspiration from MatFormer, we propose \textbf{Width-wise Starbucks}, which reduces the dimensionality of internal layers instead of the number of layers, providing a complementary axis to depth-wise scaling. Separately, we introduce \textbf{Starbucks Masked Autoencoding (SMAE)}, a Matryoshka-style pre-training strategy tailored to embedding models, inspired by masked autoencoding approaches developed for dense retrieval~\cite{gao-callan-2021-condenser,xiao-etal-2022-retromae,wang-etal-2023-simlm,zhuang2023typo,lu2021less}.

\section{Method}

The Starbucks framework enables efficient and flexible representation learning by jointly training multiple sub-models within a single transformer-based encoder. These sub-models vary in size and computational cost, and are designed to support diverse deployment-time requirements. Starbucks consists of two components: (1) Starbucks Representation Learning (SRL) for fine-tuning on downstream tasks, and (2) Starbucks Masked Autoencoding (SMAE) for pre-training the encoder.

At a high level, each sub-model in Starbucks is characterized by:
(1) A \textbf{final embedding dimension} $d$, which determines the size of the vector used for retrieval and directly affects the \emph{search time}~\footnote{During similarity search, as smaller dimension will reduce the computation required}; and (2) An \textbf{encoder model size}, which determines the cost of encoding query online and encoding passages offline. 
Figure~\ref{fig:efficiency-heatmap} in the Appendix illustrates how Starbucks-sized sub-models can significantly reduce online latency for both encoding and retrieval, compared to using the full model.
We explore two orthogonal approaches for reducing encoding cost:
(1) \textbf{Depth-wise Starbucks}, where the number of transformer layers $n$ is reduced (e.g., using only the first 2, 4, or 6 layers). This was used in the same way when the original 2D Matryoshka models were introduced~\cite{li2024eseespressosentenceembeddings}; (2) \textbf{Width-wise Starbucks}, where each layer is thinned by reducing the intermediate feed-forward dimension and attention hidden size (e.g., halving from 3072 to 1536 or 512). Note that even though cutting model width was investigated in Matformer~\cite{devvrit2023matformernestedtransformerelastic}, it was never investigated in the context of 2D Matryoshka models.

In this work, we focus primarily on the \emph{depth-wise} variant for analysis and benchmarking. The \emph{width-wise} variant is included as an ablation (Section~\ref{section:width}) to examine trade-offs under constant depth but reduced per-layer complexity.

\subsection{Starbucks Representation Learning (SRL)}

Given an input text $x$, we define a sub-model by selecting: the first $n$ layers of the encoder to process the input, and the first $d$ dimensions of the final representation.\footnote{Here we describe the formulation using examples from Depth-wise Starbucks.}

The encoder produces contextualized token embeddings:
\begin{equation}
	\begin{split}
		V_{n}^d &= \textrm{encoder}_{1:n}(x)_{1:d} \in \mathbb{R}^{|x| \times d} \\
		v_{n}^d &= \textrm{pool}(V_{n}^d) \in \mathbb{R}^{1 \times d},
	\end{split}
\end{equation}
where $\textrm{pool}(\cdot)$ denotes a pooling function (e.g., mean pooling or CLS token pooling), and $v_{n}^d$ is the resulting embedding vector.

We define a set $S = \{(n_1, d_1), \dots, (N, D)\}$, which includes a range of sub-model configurations from small to full size. For example, in our Depth-wise Starbucks using a BERT-base encoder ($N=12$, $D=768$), we set $S = \{(2, 32), (4, 64), (6, 128), (8, 256), (10, 512), (12,\\768)\}$.
The primary loss objective is:
\begin{equation}
	\mathcal{L}_{S} = \frac{1}{|S|} \sum_{(n, d)\in S} \textrm{loss}(v_{n}^d; A),
\end{equation}
where $\textrm{loss}(\cdot)$ is a task-specific contrastive or similarity-based loss function, and $A$ denotes the training batch.

To further encourage consistency across sub-models, we apply a KL divergence loss between the logits of each sub-model $(n, d)$ and the full model $(N, D)$:
\begin{equation}
	\mathcal{L}_{\textrm{KL}} = \frac{1}{|S|} \sum_{(n, d)\in S} \textrm{KLD}(s_{n}^d, s_{N}^D),
\end{equation}
where $s_{n}^d$ is the score distribution (e.g., similarity logits) produced by sub-model $(n, d)$.
The total SRL training loss is:
\[
\mathcal{L}_{\textrm{SRL}} = \mathcal{L}_S + \mathcal{L}_{\textrm{KL}}.
\]

\noindent\textbf{Depth-wise vs. Width-wise Variants.}
The \textbf{depth-wise} variant reduces the number of transformer layers $n$ while keeping layer widths unchanged.
The \textbf{width-wise} variant fixes $n = N$ and reduces intermediate and attention dimensions.
Both share a common embedding dimension $d$ for downstream retrieval, enabling fair search-time comparison (see Section~\ref{section:width}).

\subsection{Starbucks Masked Autoencoding (SMAE)}
\label{sec:method_smae}

To provide a stronger initialization for SRL, we propose \textbf{SMAE}, which adapts masked language modeling pretraining to support multiple submodels. The architecture consists of a shared encoder and a lightweight decoder.\footnote{Following RetroMAE, we adopt a one-layer decoder architecture derived from BERT.} Given an input sequence $x$, masked variants are generated as follows:
\begin{equation}
	\begin{split}
		x_{\textrm{enc}} &= \textrm{Mask}(x, p_{\textrm{enc}}), \\
		x_{\textrm{dec}} &= \textrm{Mask}(x, p_{\textrm{dec}}),
	\end{split}
\end{equation}
with $p_{\textrm{dec}} > p_{\textrm{enc}}$, so that the decoder receives less context and must rely on the encoder’s output for reconstruction, thereby reinforcing encoder training. The CLS token in $x_{\textrm{dec}}$ is replaced with the encoder’s output for a sub-model $(n, d)$:
\begin{equation}
	\begin{split}
		V_{n}^d &= \textrm{encoder}_{1:n}(x_{\textrm{enc}})_{1:d} \\
		\vec{x}_{\textrm{dec}}^{\textrm{cls}} &= \textrm{pool}(V_{n}^d \cdot W_{1:d}),
	\end{split}
\end{equation}
where $W \in \mathbb{R}^{D \times D}$ projects sub-dimension outputs to full size.
The MLM loss is computed for both encoder and decoder:
\begin{equation*}
	\begin{split}
		\mathcal{L}_{\textrm{enc}}(n, d) &= -\frac{1}{|M_{\textrm{enc}}|} \sum_{i \in M_{\textrm{enc}}} \log P(x_i | V_{n}^d) \\
		\mathcal{L}_{\textrm{dec}}(n, d) &= -\frac{1}{|M_{\textrm{dec}}|} \sum_{i \in M_{\textrm{dec}}} \log P(x_i | \vec{x}_{\textrm{dec}}^{\textrm{cls}}, x_{\textrm{dec}})
	\end{split}
\end{equation*}
and the combined pre-training loss is:
\begin{equation*}
	\mathcal{L}_{\textrm{SMAE}} = \frac{1}{|S|} \sum_{(n,d)\in S} \left[ \mathcal{L}_{\textrm{enc}}(n,d) + \mathcal{L}_{\textrm{dec}}(n,d) \right].
\end{equation*}

After SMAE pre-training, the decoder is discarded. The encoder is then used as the initialization for Starbucks SRL. The SMAE strategy is applied to both depth-wise and width-wise settings, with $S$ defined according to the corresponding sub-model architectural variant.
\section{Experiment Setup}

\subsection{Starbucks Masked Autoencoding}

We pre-train the SMAE using the BERT-base architecture~\cite{devlin-etal-2019-bert}\footnote{\href{https://huggingface.co/google-bert/bert-base-uncased}{huggingface.co/google-bert/bert-base-uncased}} on 10 billion tokens from the Fineweb dataset~\cite{penedo2024finewebdatasetsdecantingweb}\footnote{\href{https://huggingface.co/datasets/HuggingFaceFW/fineweb}{huggingface.co/datasets/HuggingFaceFW/fineweb}}.  
Training is conducted for one epoch with a batch size of 512, using AdamW (learning rate of 1e-4, weight decay 0.05), cosine learning rate decay, and a warmup ratio of 0.05. 
We implement the training process using the Hugging Face Transformers library~\cite{wolf-etal-2020-transformers}. The pre-trained SMAE checkpoint is available at \href{https://huggingface.co/ielabgroup/bert-base-uncased-fineweb100bt-smae}{ielabgroup/bert-base-uncased-fineweb100bt-smae}.

\subsection{Fine-tuning and Evaluation}

We organize our evaluation into two parts: \textbf{in-domain evaluation}, which assesses performance on tasks aligned with the training data, and \textbf{out-of-domain generalization}, which tests robustness across diverse unseen domains.

\noindent\textbf{In-domain Evaluation.}
We evaluate Starbucks on two downstream tasks: \textbf{semantic text similarity} (STS) and \textbf{passage retrieval}. Fine-tuning is performed using our SRL tuning approach and the 2DMSE baseline, starting from either the original BERT or our SMAE-pretrained models.
For \textbf{STS}, we follow standard practice by training on the MultiNLI~\cite{williams-etal-2018-broad} and SNLI~\cite{bowman-etal-2015-large} datasets, and evaluating on STS 2012–2016~\cite{agirre-etal-2016-semeval}, SICK-R~\cite{marelli-etal-2014-sick}, and STS-B~\cite{cer-etal-2017-semeval}. Evaluation is based on Spearman’s correlation. We use the Sentence Transformers library~\cite{reimers-gurevych-2019-sentence}, extended to support SRL-style fine-tuning.
For \textbf{passage retrieval}, we train on the MS MARCO Passage Ranking dataset~\cite{bajaj2018msmarcohumangenerated} using Tevatron~\cite{tevatron}. Evaluation is conducted on the MS MARCO development set using MRR@10, and on the TREC DL19 and DL20 benchmarks~\cite{craswell2020overviewtrec2019deep, craswell2021overviewtrec2020deep} using nDCG@10. These benchmarks reflect both binary and graded relevance and are considered in-domain. The hyperparameters used for fine-tuning are summarized in Appendix~\ref{appendix:parameters}.

\begin{table*}[htp!]
	\centering
	\scriptsize
	\caption{In-domain evaluation on STS and Passage Retrieval, using averaged metrics across all Starbucks sizes: Spearman’s correlation for STS, MRR@10 for MS MARCO dev, and NDCG@10 for DL19/DL20. }
	\label{tab:results}
	\begin{tabular}{l|cccccccc|ccc}
		\toprule
		&\multicolumn{8}{c|}{\textbf{Semantic Text Similarity Tasks}} &\multicolumn{3}{c}{\textbf{Retrieval Tasks}} \\ \midrule
		\textbf{Method} & \textbf{STS-B} & \textbf{STS12} & \textbf{STS13} & \textbf{STS14} & \textbf{STS15} & \textbf{STS16} & \textbf{SICK-R} & \textbf{Average} & \textbf{MARCO dev}& \textbf{DL19} & \textbf{DL20} \\
		\midrule
		BERT & 0.6707 & 0.5525 & 0.6607 & 0.6036 & 0.7219 & 0.6902 & 0.6656 & 0.6522 & 0.0559 & 0.1084
		 & 0.1076 \\
		\midrule
		BERT-Separate & 0.7872 & 0.7128 & 0.7742 & 0.7269 & 0.8075 & 0.7722 & 0.7702 & 0.7644 & 0.2917 & 0.5859 & 0.5857\\
		\midrule
		BERT-2DMSE & 0.7629 & 0.6691 & 0.7282 & 0.6793 & 0.7883 & 0.7501 & 0.7584 & 0.7338 & 0.2560 & 0.5218 &  0.5408\\ 
		BERT-SRL  &0.7906 & 0.7197 & 0.7742 & 0.7252 &0.8126& 0.7784 & \textbf{0.7766} & 0.7682 & 0.2921 & 0.6009 & \textbf{0.6126} \\ \midrule
		
		
		
		SMAE-SRL~\includegraphics[height=8pt]{figures/icon.pdf}&  \textbf{0.8073} & \textbf{0.7286} & \textbf{0.7868} & \textbf{0.7558} & \textbf{0.8362} & \textbf{0.7948} & 0.7764 & \textbf{0.7837} & \textbf{0.3116} & \textbf{0.6135} & 0.6039 \\
		\midrule
	\end{tabular}
\end{table*}

\noindent\textbf{Out-of-domain Generalization.}
To assess generalization, we evaluate models fine-tuned on MS MARCO across 13 BEIR datasets, following standard practice~\cite{beir}.\footnote{We exclude three private BEIR datasets, CQADupStack (due to evaluation issues), and MS MARCO dev (in-domain).} This setup tests Starbucks' ability to generalize to diverse domains in a zero-shot retrieval setting.

\section{Results}

Results are presented in two parts: in-domain (Section~\ref{sec:results_in_domain}), where test data match the training distribution, and out-of-domain (Section~\ref{sec:ood_results}), which evaluates generalization to new retrieval domains.

\subsection{In-domain Evaluation}
\label{sec:results_in_domain}

Table~\ref{tab:results} presents the average effectiveness across six Starbucks sizes, $S = \{(2, 32), (4, 64), \ldots, (12, 768)\}$, on STS and passage retrieval (Per-size results are presented at Appendix in Table~\ref{table:full}). We compare several training strategies: (i) full-size BERT models fine-tuned on each task and then sliced for evaluation (BERT), (ii) one BERT model fine-tuned separately for each Starbucks configuration (BERT-Separate), (iii) the 2D Matryoshka training method (BERT-2DMSE)~\footnote{We use 2DMSE-v1 due to ongoing updates to the model.}, (iv) our SRL fine-tuning method (BERT-SRL), and (v) the full Starbucks approach combining SMAE pre-training and SRL fine-tuning (SMAE-SRL).

Overall, Starbucks outperforms all baselines on average, with the only exceptions occurring on SICK-R (STS) and DL20 (retrieval). SRL alone (BERT-SRL) already provides consistent improvements over 2DMSE and separate models across most datasets. These gains are further amplified when combined with SMAE pre-training.

\begin{figure}[t]
	\centering
	\tiny
	\begin{subfigure}[b]{0.48\linewidth}
		\includegraphics[width=\linewidth]{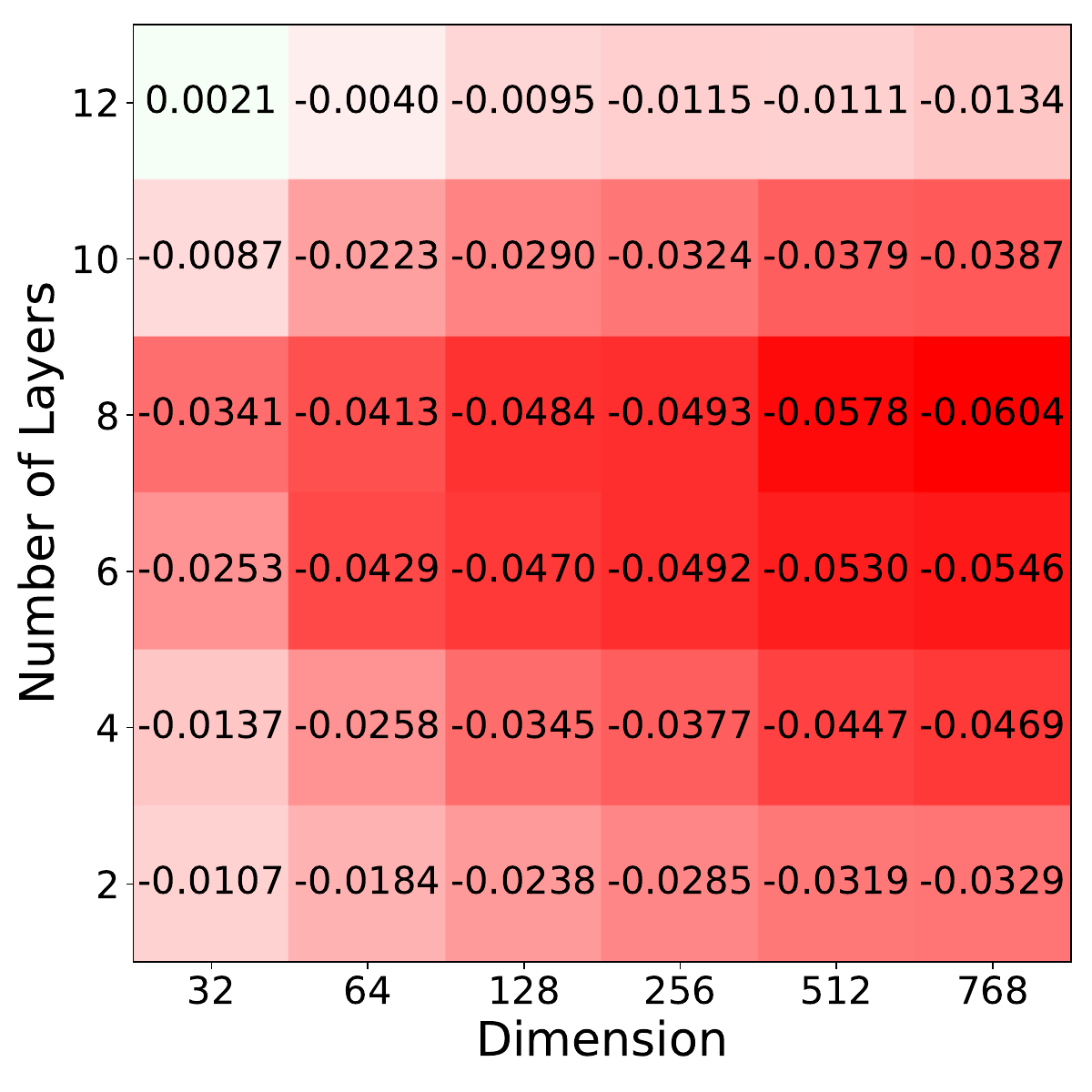}
		\caption{STS Average}
		\label{fig:small_scale_vs_2dmse}
	\end{subfigure}
	\begin{subfigure}[b]{0.48\linewidth}
		\includegraphics[width=\linewidth]{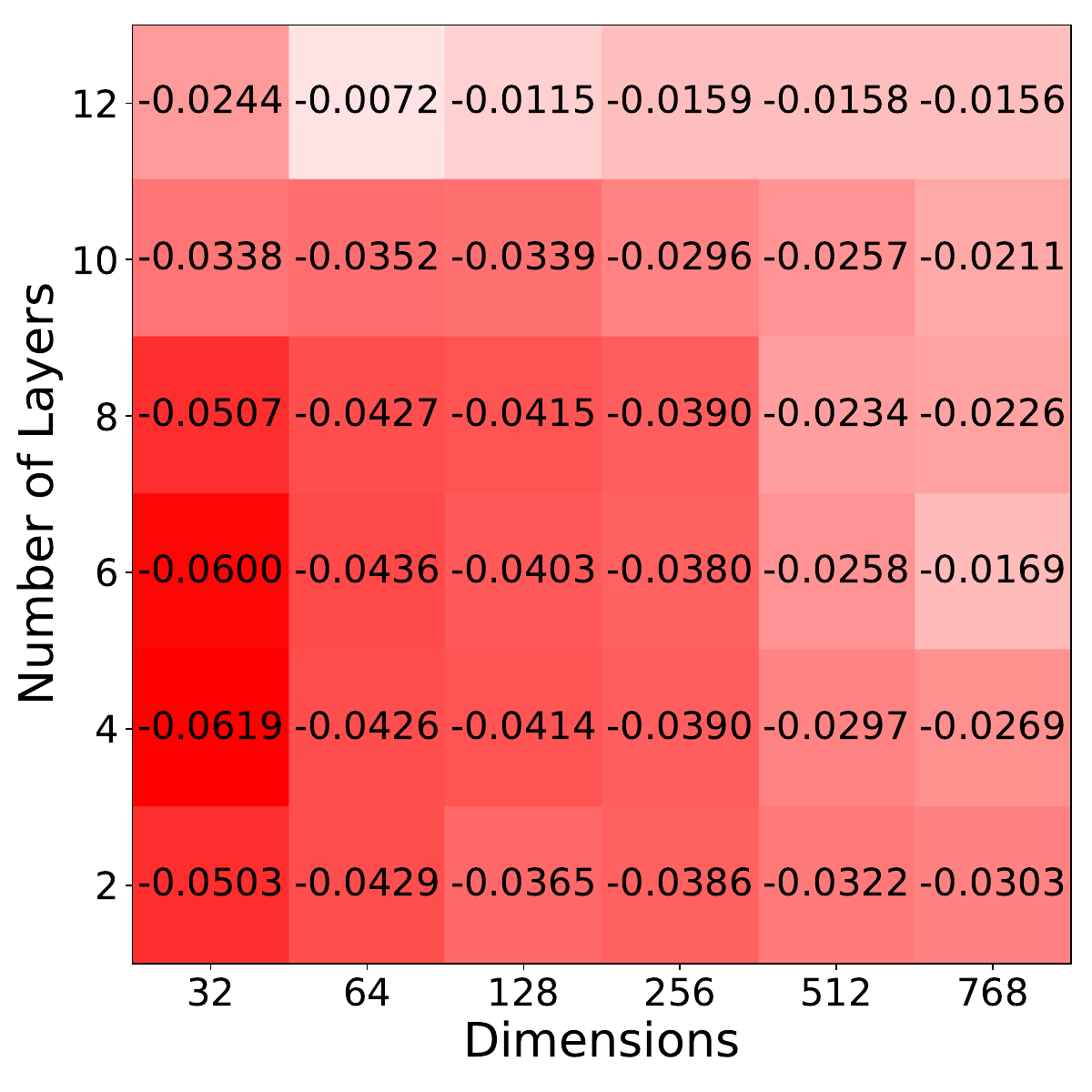}
		\caption{MARCO dev}
		\label{fig:msmarco_small_scale_vs_2dmse}
	\end{subfigure}
\caption{
	Comparison of 2DMSE and separately trained models on (a) STS (Spearman's correlation) and (b) MS MARCO dev (MRR@10).Negative values (red) indicate  separately trained models are better.}
	\label{fig:comparison_2dmse_vs_separate}
\end{figure}

\subsubsection{2DMSE vs. Separate Training}
\label{sec:small_scale_vs_2dmse}

Figure~\ref{fig:comparison_2dmse_vs_separate} shows the performance difference between 2DMSE and separately trained models.
 We find that separate training consistently yields better results across all configurations, with the exception of layer 12 and dimension 32. The gap is particularly pronounced for shallow layers, indicating that 2DMSE compromises effectiveness when distributing training across many subnetworks. 
 Furthermore, We observe that layer depth has a greater impact on effectiveness than embedding dimension. A possible explanation lies in the original 2DMSE loss setup, where only one early layer and the final layer are used—giving the last layer disproportionate weight. This may weaken supervision for earlier layers\footnote{We also tried adding all layers in the loss, which slightly improved early layers but significantly hurt final-layer performance.}.

\begin{figure}[t]
	\centering
	\tiny

	\begin{subfigure}[b]{0.48\linewidth}
		\includegraphics[width=\linewidth]{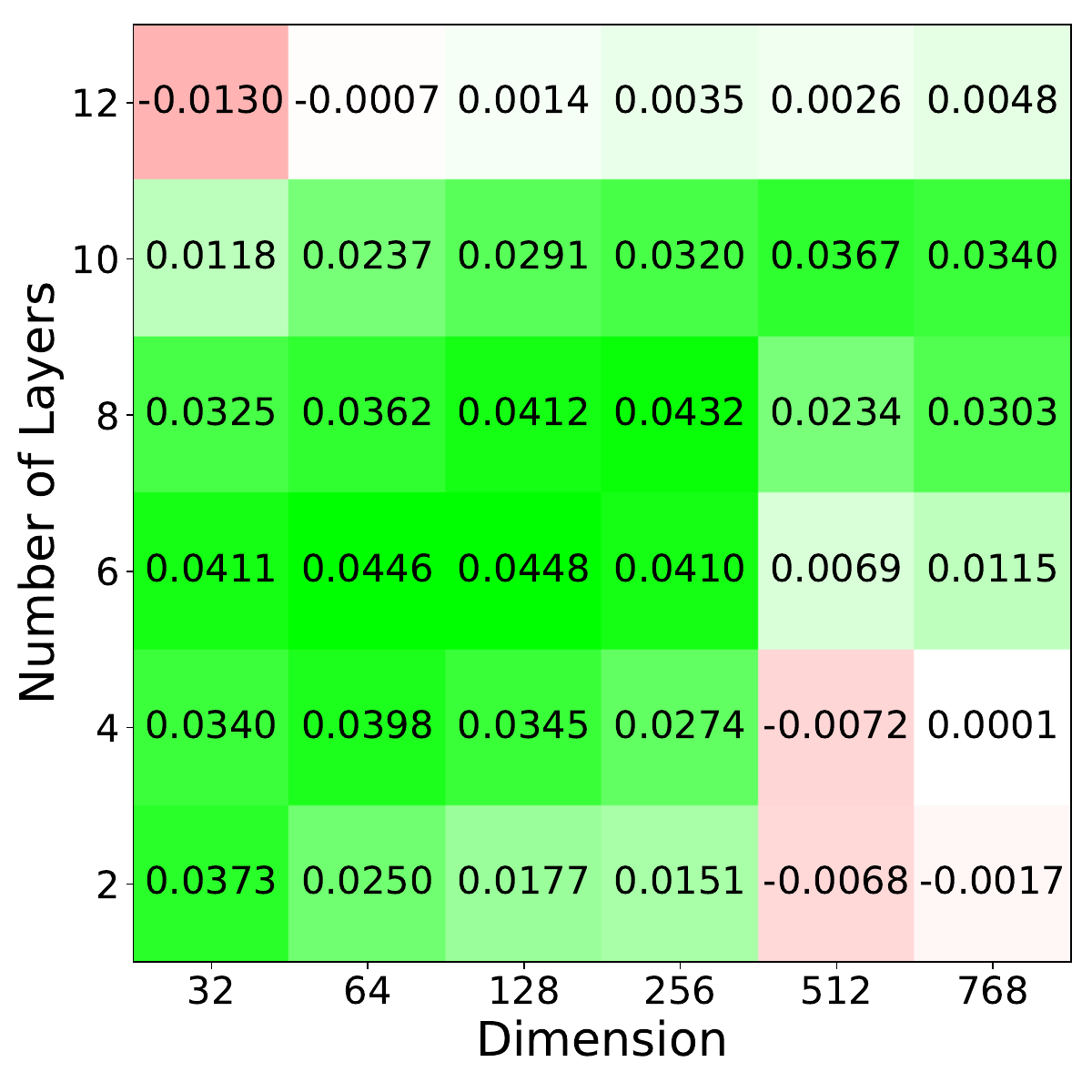}
		\caption{STS Average}
	\end{subfigure}
	\begin{subfigure}[b]{0.48\linewidth}
		\includegraphics[width=\linewidth]{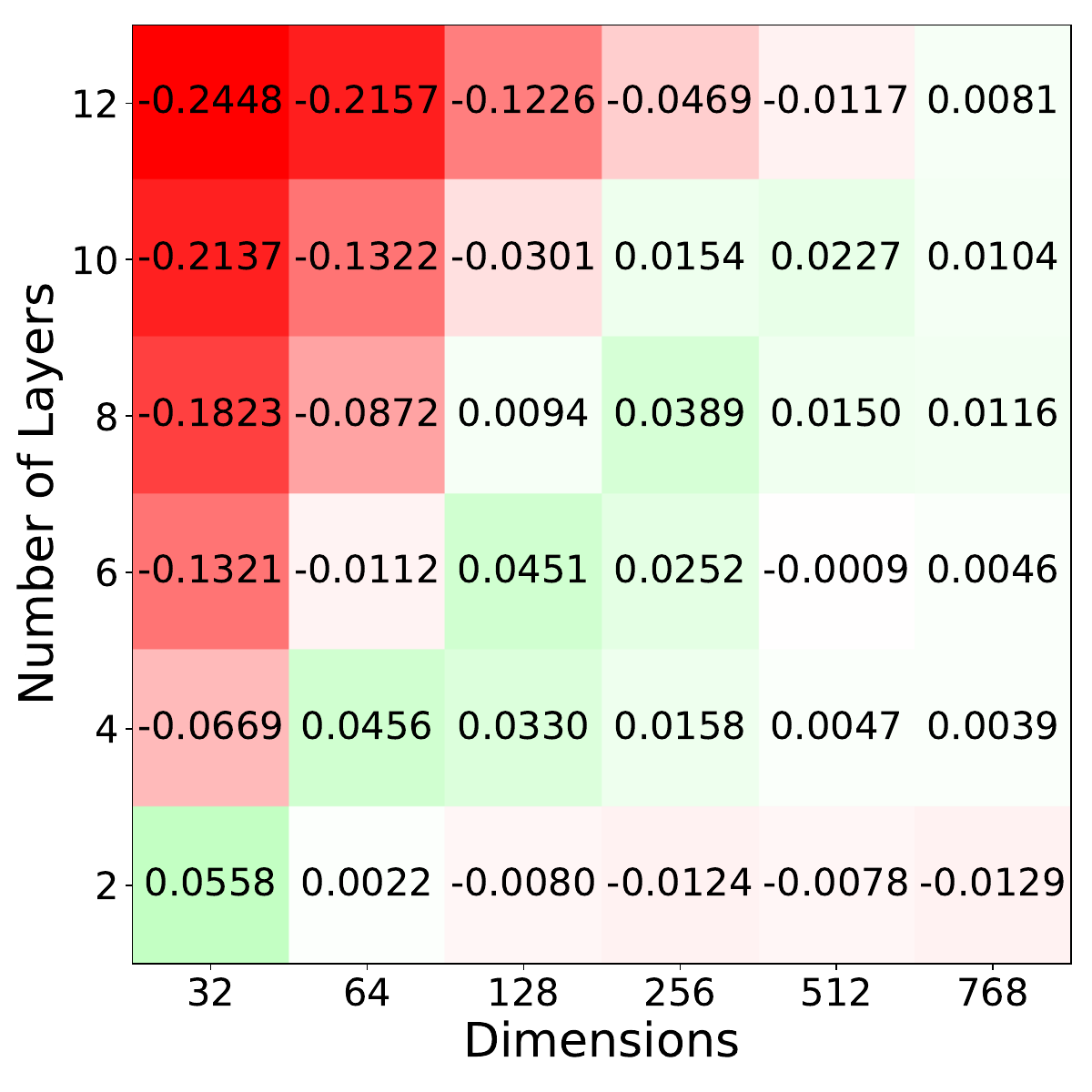}
		\caption{MARCO dev}
		\label{fig:msmarco_srl_vs_2d}
	\end{subfigure}
	\caption{
		Comparison of SRL and 2DMSE on (a) STS (Spearman's correlation) and (b) MS MARCO dev (MRR@10). Positive values (green) indicate SRL is better; negative values (red) indicate 2DMSE is better.}
	\label{fig:comparison_plots_finetune}
\end{figure}

\begin{figure}[t]
	\centering
	\tiny
	\begin{subfigure}[b]{0.48\linewidth}
		\includegraphics[width=\linewidth]{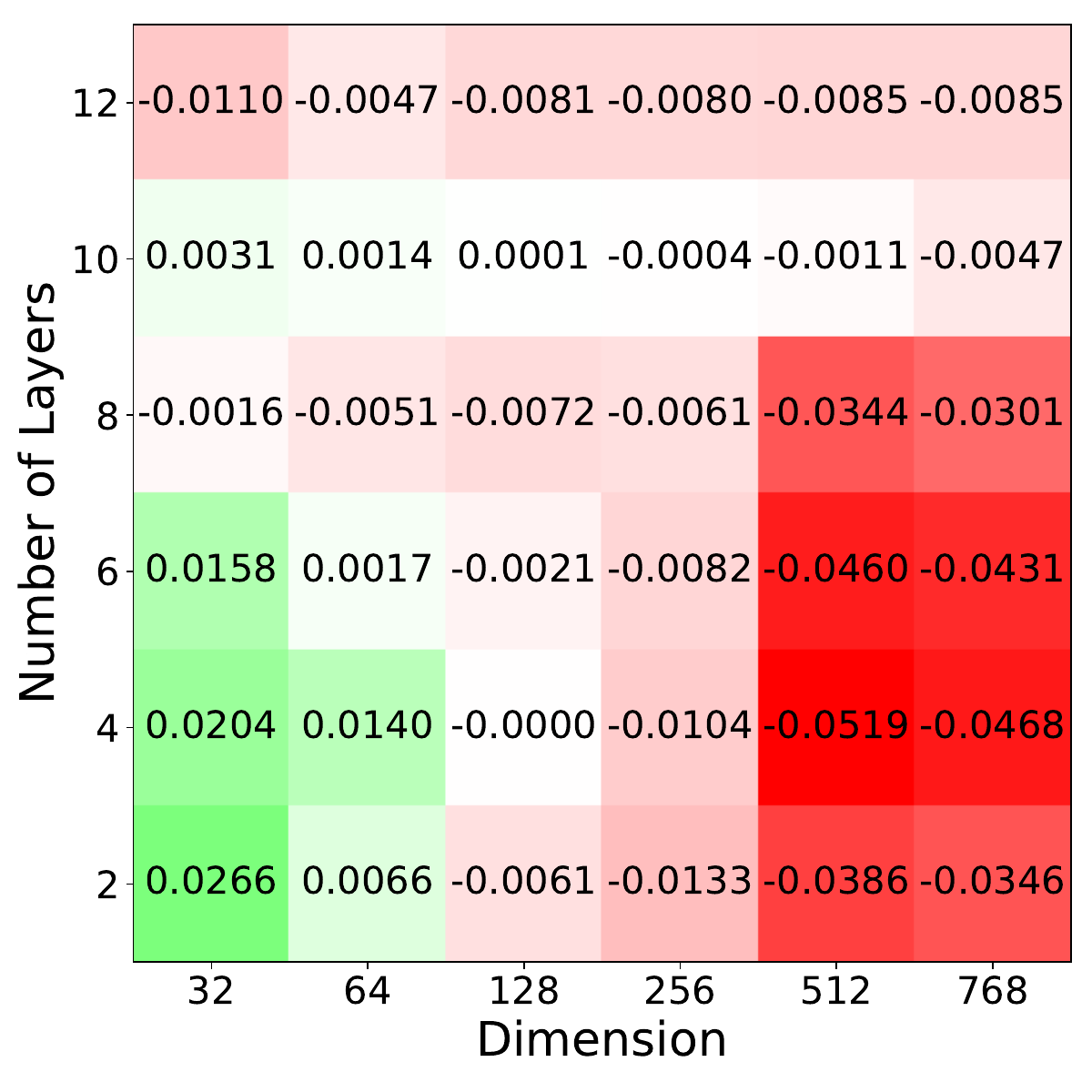}
		\caption{STS Average}
		\label{fig:sts_srl_vs_baseline}
	\end{subfigure}
	\hfill
	\begin{subfigure}[b]{0.48\linewidth}
		\includegraphics[width=\linewidth]{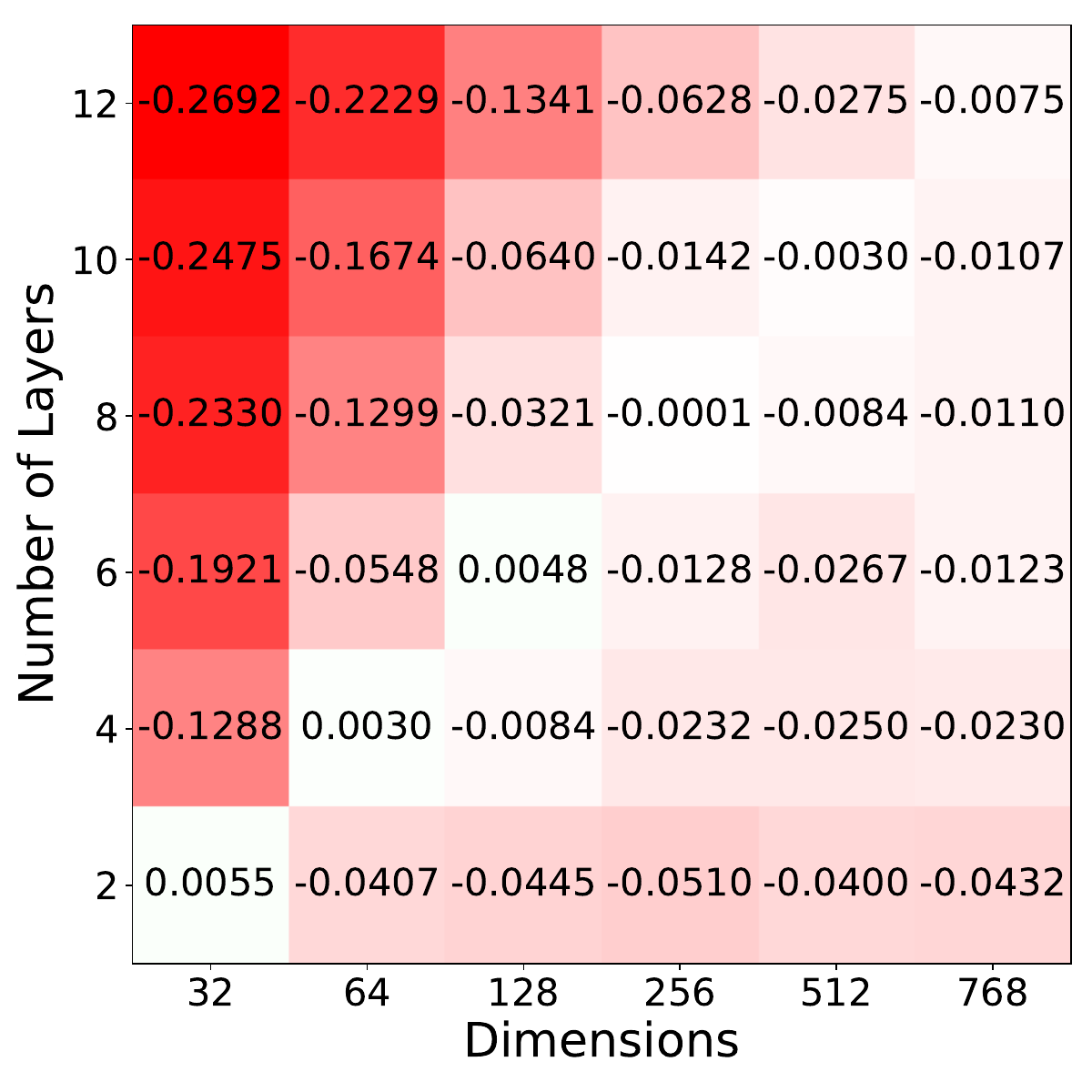}
		\caption{MARCO dev}
		\label{fig:msmarco_srl_vs_baseline}
	\end{subfigure}
\caption{
	Comparison of SRL and separately trained models on (a) STS (Spearman's correlation) and (b) MS MARCO dev(MRR@10). Positive values (green) indicate SRL is better; negative values (red) indicate seperately trained model is better.  }
	\label{fig:comparison_plots_baseline}
\end{figure}

\subsubsection{SRL vs. Baselines}
\label{sec:SRL_vs_baselines}

Figure~\ref{fig:comparison_plots_finetune} compares SRL with both 2DMSE and separately trained models. SRL fine-tuning is performed in a single training run from the original BERT checkpoint, targeting six predefined Starbucks sizes. Despite this constraint, SRL generalizes well to nearby configurations and consistently outperforms 2DMSE on the STS task. We hypothesize that this is due to the complexity of the 2DMSE training objective, which may lead to suboptimal optimization and degraded performance across all configurations.

The most substantial gains are observed along the Starbucks diagonal—that is, when the number of layers and embedding dimensions match those seen during training—highlighting the benefit of explicitly training for these size pairs.

Figure~\ref{fig:comparison_plots_baseline} presents a comparison between SRL and upper-bound separately trained models. Here, SRL achieves comparable, and in some cases higher effectiveness, particularly in smaller configurations with shallow depths. Performance on off-diagonal configurations (e.g., 12 layers with 32 dimensions) tends to be weaker, likely because SRL’s training objective emphasizes the diagonal pairs, resulting in less effective generalization to mismatched sizes.

\begin{figure}[t]
	\centering
	\tiny
	\begin{subfigure}[b]{0.48\linewidth}
		\includegraphics[width=\linewidth]{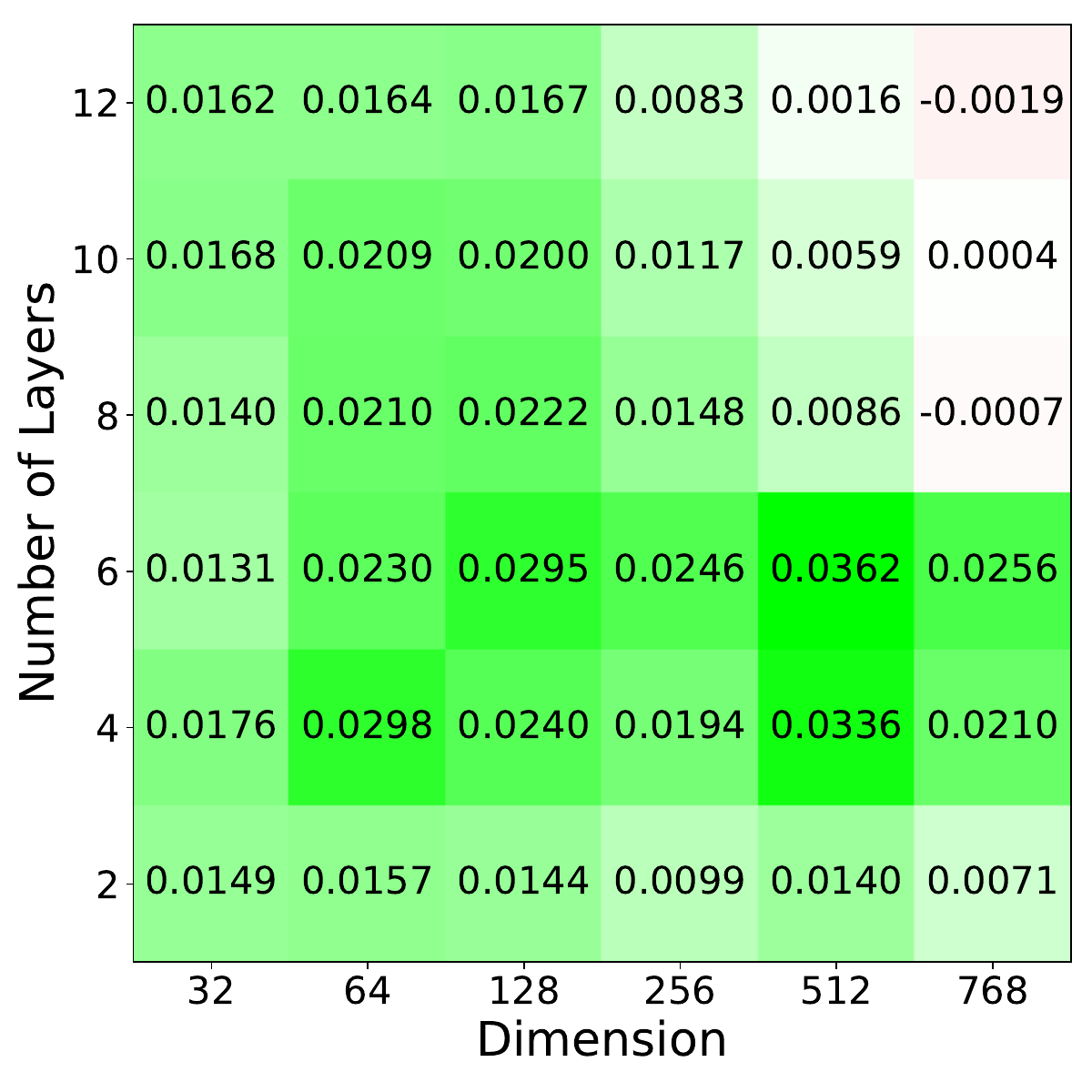}
		\caption{STS}
		\label{fig:starbucks_vs_srl_sts}
	\end{subfigure}
	\hfill
	\begin{subfigure}[b]{0.48\linewidth}
		\includegraphics[width=\linewidth]{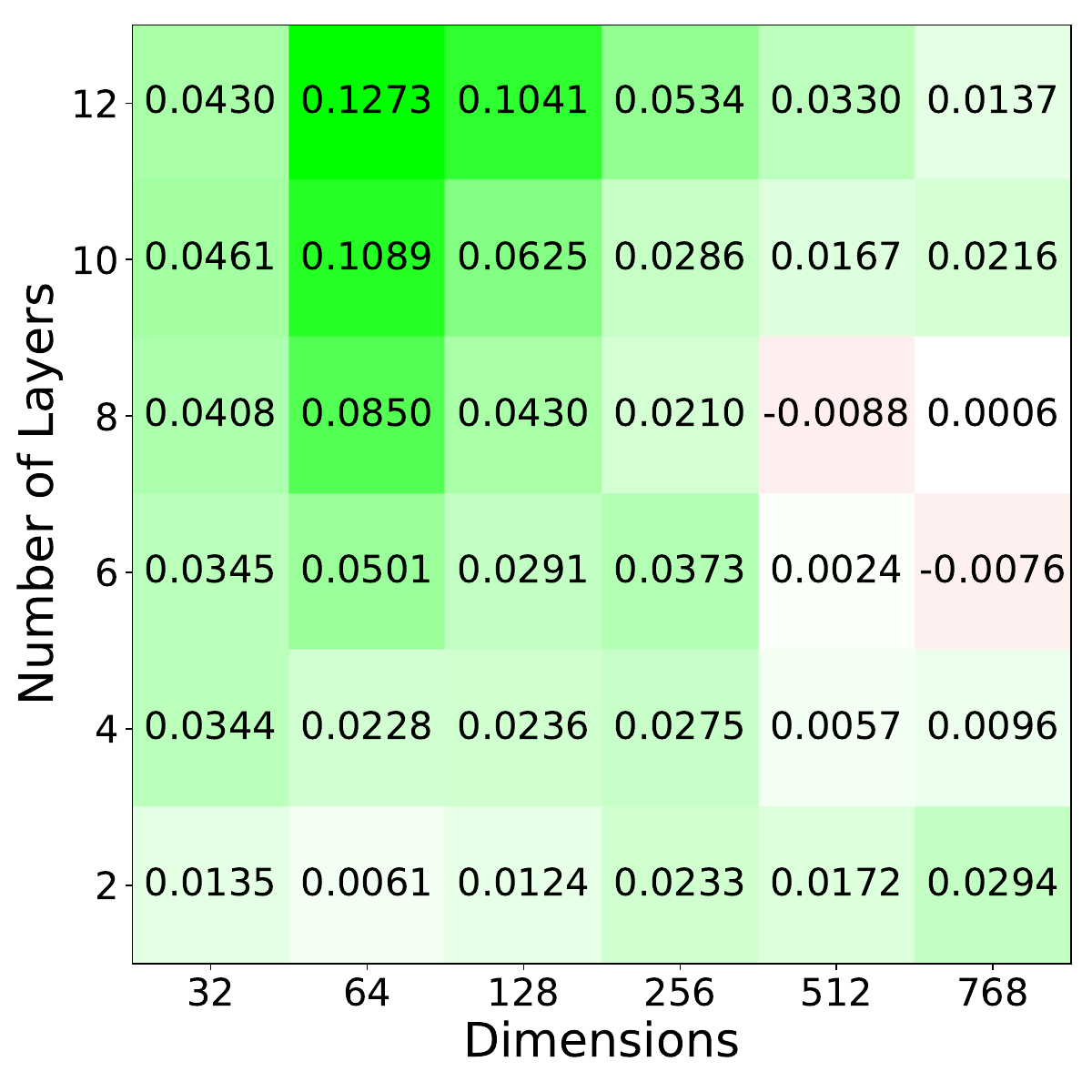}
		\caption{MS MARCO}
		\label{fig:starbucks_vs_srl_msmarco}
	\end{subfigure}
	\caption{
		Comparison of SMAE-SRL and BERT-SRL on (a) STS (Spearman's correlation) and (b) MS MARCO dev (MRR@10). Positive values (green) indicate better performance by SMAE-SRL.}
	\label{fig:starbucks_vs_srl_combined}
\end{figure}

\subsubsection{Impact of SMAE Pre-training}
\label{sec:smae}

Figure~\ref{fig:starbucks_vs_srl_combined} presents a comparison between models fine-tuned with SRL using different initialization strategies: BERT versus SMAE. The results show that SMAE pre-training consistently outperforms the BERT baseline, particularly for smaller configurations (i.e., fewer layers and lower-dimensional embeddings). The only exception is in full-size models, where performance is marginally lower—potentially explaining why certain datasets, such as SICK-R, show slightly reduced effectiveness under SMAE. These findings suggest that SMAE, which is designed to learn compressed representations, is particularly well-suited for initializing sub-networks that are subsequently fine-tuned with SRL.

\begin{table*}[htp!]
	\centering
	\scriptsize
	\caption{Out-of-domain evaluation on BEIR datasets, represented by averaged NDCG@10 on all Starbuck sizes.}
	\label{tab:beir_results}
	\resizebox{\linewidth}{!}{
		\begin{tabular}{@{}l|cc|ccc|cc|c|cc|cc|c@{}}
			\toprule
			& \multicolumn{2}{c|}{\textbf{General}} & \multicolumn{3}{c|}{\textbf{Scientific}} & \multicolumn{2}{c|}{\textbf{Debate}} & \multicolumn{1}{c|}{\textbf{Finance}} & \multicolumn{2}{c|}{\textbf{Biomedical}} & \multicolumn{2}{c|}{\textbf{Wiki}} &  \\
			\midrule
			\textbf{Method} & \textbf{Db.\-Ent.} & \textbf{Quora} & \textbf{Scidocs} & \textbf{Scifact} & \textbf{Cli.\-Fever} & \textbf{Arguana} & \textbf{Touche20} & \textbf{Fiqa} & \textbf{Trec\-Cvd} & \textbf{Nfcorpus} & \textbf{Fever} & \textbf{Hpqa} & \textbf{Avg.} \\
			\midrule
			BERT & 0.0511 & 0.2355 & 0.0203 & 0.0918 & 0.0281 & 0.0553 & 0.0389 & 0.0376 & 0.1140 & 0.0605 & 0.1101 & 0.0761 & 0.0766 \\
			\midrule
			BERT-Separate & 0.2551 & \textbf{0.7816} & 0.0894 & 0.4087 & 0.1160 & 0.2269 & 0.2113 & 0.1760 & 0.5362 & 0.2246 & \textbf{0.5655} & 0.3634 & 0.3296 \\ \midrule
			BERT-2DMSE & 0.2079 & 0.5849 & 0.0691 & 0.3300 & 0.1057 & 0.1991 & 0.2049 & 0.1392 & 0.4762 & 0.1871 & 0.4588 & 0.2611 & 0.2687 \\
			BERT-SRL & 0.2483 & 0.6994 & 0.0888 & 0.4083 & 0.1215 & 0.2222 & 0.2249 & 0.1702 & 0.5458 & 0.2202 & 0.5489 & 0.3543 & 0.3211 \\
			\midrule
			SMAE-SRL \includegraphics[height=8pt]{figures/icon.pdf}& \textbf{0.2610} & 0.6559 & \textbf{0.1011} & \textbf{0.4131} & \textbf{0.1327} & \textbf{0.2445} & \textbf{0.2329} & \textbf{0.2096} & \textbf{0.6190} & \textbf{0.2473} & 0.5619 & \textbf{0.3708} & \textbf{0.3375} \\
			\bottomrule
		\end{tabular}
	}
\end{table*}

\subsection{Out-of-domain Evaluation}
\label{sec:ood_results}

To evaluate the generalization ability of Starbucks, we assess Starbucks variants and baseline models in a zero-shot setting: all models are fine-tuned on MS MARCO and directly evaluated—without further adaptation—on 13 datasets from the BEIR benchmark. Average results are shown in Table~\ref{tab:beir_results}, with full per-Starbuck size results in Table~\ref{tab:beir_sizes} (Appendix).

The results show that SRL matches the effectiveness of BERT models trained separately for each size. Notably, SRL requires only a single training run and supports multiple configurations at inference by adjusting layer-dimension parameters.

With SMAE pre-training, SRL outperforms separately trained BERT variants on 11 of 13 datasets, demonstrating strong improvements in zero-shot generalization. These results confirm that the gains observed in in-domain settings transfer robustly to diverse retrieval tasks in BEIR.

\begin{table*}[!t]\centering
	\caption{Impact of KL divergence loss $\mathcal{L}_{\textrm{KL}}$. STAB Spearman's Correlation is used for STS task, and MARCO dev MRR@10 is used for Retrieval.}\label{tab:kl_div}
	\scriptsize
	\resizebox{0.99\linewidth}{!}{
	\begin{tabular}{llrrrrrrr}\toprule
		&&Demi (2, 32) &Short (4, 64) &Tall (6, 128) &Grande (8, 256) &Venti (10. 512) &Trenta (12, 768) &Average \\\midrule
		\multirow{2}{*}{STS} &	w. $\mathcal{L}_{\textrm{KL}}$ & 0.7455 & \textbf{0.7933} & \textbf{0.8202} & \textbf{0.8278} & \textbf{0.8292} & \textbf{0.8274} & \textbf{0.8073} \\
		&w/o. $\mathcal{L}_{\textrm{KL}}$ & \textbf{0.7463} & 0.7925 & 0.8190 & 0.8274 & 0.8283 & 0.8269 & 0.8067 \\
	\midrule

		\multirow{2}{*}{Retrieval} & w. $\mathcal{L}_{\textrm{KL}}$&\textbf{0.2282} &\textbf{0.2950} &\textbf{0.3274} &\textbf{0.3369} &\textbf{0.3416} &0.3403 &\textbf{0.3116} \\
		&w/o. $\mathcal{L}_{\textrm{KL}}$ &0.2205&0.2885&0.3162&0.3295&0.3372&\textbf{0.3405}&0.3054 \\\hline
		
	\end{tabular}
}
\end{table*}

\section{Impact of Starbucks Modeling Choices}

In this section, we investigate variations in modeling choices and their impact on embedding quality. We focus on three key ablation studies. First, we examine the role of \textbf{KL divergence} during fine-tuning and its influence on model performance (Section~\ref{sec:kl_importance}). Second, we evaluate the effect of different \textbf{pre-training strategies} (Section~\ref{sec:pretraining_method}). Finally, we test the applicability of the SRL fine-tuning approach to a recent \textbf{state-of-the-art embedding backbone}, E5 (Section~\ref{sec:e5}).

For these ablation studies, we evaluate on a single dataset per task: STAB for semantic textual similarity (STS), and MS MARCO (dev) for passage retrieval.

\begin{table*}[t]\centering
	\caption{Impact of pre-training components. STAB Spearman's Correlation is used for the STS task, and MARCO dev MRR@10 is used for the retrieval task.}\label{tab:pretraining}
	\scriptsize
	\resizebox{\linewidth}{!}{
	\begin{tabular}{llrrrrrrr}\toprule
		&&Demi (2, 32) &Short (4, 64) &Tall (6, 128) &Grande (8, 256) &Venti (10. 512) &Trenta (12, 768) &Average \\\midrule
		
		\multirow{3}{*}{STS} &	SMAE& 0.7455 & 0.7933 & 0.8202 & \textbf{0.8278} & \textbf{0.8292} & 0.8274 & \textbf{0.8073} \\
		&- MAE  & \textbf{0.7475} & \textbf{0.7968} & \textbf{0.8211} & 0.8257 & 0.8288 & 0.8239 & \textbf{0.8073} \\
		&- Starbucks& 0.7399 & 0.7675 & 0.7902 & 0.8134 & 0.8238 & \textbf{0.8360} & 0.7951   \\\midrule

		\multirow{3}{*}{Retrieval} &SMAE &0.2281 &0.2949 &\textbf{0.3273} &\textbf{0.3369} &\textbf{0.3416} &\textbf{0.3403} &\textbf{0.3116} \\
		&- MAE &\textbf{0.2349} &\textbf{0.2972} &0.3237 &0.3282 &0.3339 &0.335 &0.3088 \\
		&- Starbucks &0.2255 &0.2859 &0.3098 &0.3277 &0.3375 &\textbf{0.3403} &0.3044 \\\hline
	\end{tabular}
}
\end{table*}

\begin{table*}[t]\centering
	\caption{Impact of pre-training backbone. STAB Spearman's Correlation is used for the STS task, and MARCO dev MRR@10 is used for the retrieval task.}\label{tab:backbone}
	\scriptsize
	\resizebox{\linewidth}{!}{
	\begin{tabular}{llrrrrrrr}\toprule
		&&Demi (2, 32) &Short (4, 64) &Tall (6, 128) &Grande (8, 256) &Venti (10, 512) &Trenta (12, 768) &Average \\\midrule
		
		\multirow{5}{*}{STS} 
		& E5-base-v2       & 0.5893 & 0.5616 & 0.5778 & 0.5842 & 0.6910 & 0.8548 & 0.6431 \\
		& BERT-SMAE-SRL    & \textbf{0.7455} & 0.7933 & \textbf{0.8202} & 0.8278 & 0.8292 & 0.8274 & 0.8073 \\ \cmidrule{2-9}
		
		& E5-2DMSE         & 0.7050 & 0.7341 & 0.7607 & 0.7871 & 0.8072 & \textbf{0.8673} & 0.7769\\
		& E5-SRL           & 0.7313 & 0.7728 & 0.7984 & 0.8243 & \textbf{0.8452} & 0.8627 & 0.8058 \\ 
		& E5-SMAE-SRL      & 0.7437 & \textbf{0.7922} & 0.8175 & \textbf{0.8302} & 0.8337 & 0.8331 & \textbf{0.8084} \\ 
		
		\midrule
		
		\multirow{5}{*}{Retrieval} 
		& E5-base-v2       & 0.0015 & 0.0021 & 0.0042 & 0.0088 & 0.0097 & \textbf{0.3606} & 0.0645\\
		& BERT-SMAE-SRL    & 0.2282 & 0.2950 & \textbf{0.3274} &  \textbf{0.3369} & 0.3416 & 0.3403 & \textbf{0.3116} \\ \cmidrule{2-9}
		
		& E5-2DMSE         & 0.1330 & 0.2062 & 0.2429 & 0.2629 & 0.2905 & 0.2953 & 0.2385  \\
		& E5-SRL           & 0.2159 & 0.2793 & 0.3115 &0.3336 & \textbf{0.3425} & 0.3503 & 0.3055  \\ 
		& E5-SMAE-SRL      & \textbf{0.2309} & \textbf{0.3013} & 0.3217 & 0.3333 &  0.3364  & 0.3369  & 0.3101  \\ 
		
		\midrule
	\end{tabular}
}
\end{table*}

\subsection{Impact of KL Divergence}
\label{sec:kl_importance}

We examine the effect of the KL divergence component in the SRL loss, defined as $\mathcal{L} = \mathcal{L}_S + \mathcal{L}_{\textrm{KL}}$, which encourages alignment between the full-size model embeddings and those of smaller layer-dimension configurations.

Results in Table~\ref{tab:kl_div} show that incorporating $\mathcal{L}_{\textrm{KL}}$ generally benefits performance across most Starbucks sizes. Notable exceptions include the Demi size on the STS task and the Trenta size on the retrieval task. Overall, the impact of $\mathcal{L}_{\textrm{KL}}$ is more substantial in the retrieval setting, while STS shows only marginal improvements.

\subsection{Impact of Pre-training Approaches}
\label{sec:pretraining_method}

To understand the effect of various pre-training strategies, we ablate components of the Starbucks pre-training method (Section~\ref{sec:method_smae}) while keeping SRL fine-tuning fixed. We compare three pre-training configurations: (1) full Starbucks with Masked Autoencoding (MAE), (2) Starbucks without MAE, and (3) the original BERT model trained only with MLM.

As shown in Table~\ref{tab:pretraining}, full Starbucks pre-training improves effectiveness across most model sizes, with the exception of Trenta, which already aligns with the full-size training objective. Smaller models (e.g., Demi, Short) sometimes perform better without the MAE component, while larger ones (e.g., Grande, Venti, Trenta) benefit more from its inclusion. These effects are more pronounced in retrieval than in STS, likely because retrieval requires finer-grained representations to match queries against large corpora, whereas STS is a simpler sentence-pair task.

\subsection{Applicability to SOTA Backbone}
\label{sec:e5}

To evaluate the generality of our SRL fine-tuning approach, we apply it to E5~\cite{wang2022text}, a recent state-of-the-art (SOTA) embedding model. Table~\ref{tab:backbone} presents the results of applying SRL and SMAE to the E5 backbone. We observe that SRL remains effective even when used with more recent and highly optimized models. In some configurations—particularly smaller model variants—SRL even improves overall effectiveness.

However, we note a slight decrease in performance for the full model. We hypothesize that this drop is due to the full E5  model already being extensively tuned across a variety of datasets; applying SRL may shift the optimization landscape in a way that prioritizes consistency across sub-networks at the cost of peak full-model retrieval performance.

Notably, combining SRL with SMAE pre-training encourages stronger representation learning in earlier layers. This sometimes comes at the expense of later-layer expressiveness, but it reinforces that even standalone SRL is sufficient when applied to strong modern backbones. These findings suggest that Starbucks strategies can be effectively applied to SOTA architectures like E5, without requiring significant changes to their pre-existing optimization pipelines.

\section{Depth-wise vs. Width-wise Starbucks}
\label{section:width}

\begin{table}[t]\centering
	\caption{Comparison of Depth-wise vs.\ Width-wise Starbucks (SRL and SMAE-SRL) on MARCO dev, measured using MRR@10. Percentage gains of \textit{Hybrid} over the stronger of \textit{Depth} or \textit{Width} are shown in parentheses. * indicates that \textit{Width} yields a significantly different result from \textit{Depth} based on a paired t-test ($p < 0.05$).}
	\scriptsize
	\label{tab:width-wise-results}
	\resizebox{\linewidth}{!}{
		\begin{tabular}{@{}l|ccc|ccc@{}}
			\toprule
			\textbf{Model} & \multicolumn{3}{c|}{\textbf{SRL}} & \multicolumn{3}{c}{\textbf{SMAE-SRL}} \\ \midrule
			& Depth & Width & Hybrid & Depth & Width & Hybrid \\
			\midrule
			Demi   & 0.215 & 0.209 & 0.253 \textcolor{gray}{(17.8\%)} & 0.228 & 0.254* & 0.277 \textcolor{gray}{(9.0\%)} \\
			Short  & 0.272 & 0.276* & 0.299 \textcolor{gray}{(8.4\%)}  & 0.295 & 0.314* & 0.324 \textcolor{gray}{(3.2\%)} \\
			Tall   & 0.298 & 0.298 & 0.319 \textcolor{gray}{(6.9\%)}  & 0.327 & 0.333 & 0.344 \textcolor{gray}{(3.2\%)} \\
			Grande & 0.316 & 0.311 & 0.330 \textcolor{gray}{(4.5\%)}  & 0.337 & 0.340 & 0.351 \textcolor{gray}{(3.1\%)} \\
			Venti  & 0.325 & 0.323 & 0.338 \textcolor{gray}{(4.0\%)}  & 0.342 & 0.344 & 0.354 \textcolor{gray}{(3.1\%)} \\
			Trenta & 0.327 & 0.327 & 0.339 \textcolor{gray}{(3.6\%)}  & 0.340 & 0.342 & 0.353 \textcolor{gray}{(3.5\%)} \\
			\midrule
			Avg. & 0.292 & 0.291 & \textbf{0.313} \textcolor{gray}{(7.1\%)} 
			& 0.312 & 0.321 & \textbf{0.334} \textcolor{gray}{(4.0\%)} \\
			\bottomrule
		\end{tabular}
	}
\end{table}

In this section, we investigate \textbf{Width-wise Starbucks}, a variant that reduces the dimensionality of the intermediate and attention layers within each transformer block, while preserving the full layer depth. This approach draws inspiration from MatFormer~\cite{devvrit2023matformernestedtransformerelastic}, and contrasts with \textbf{Depth-wise Starbucks}, examined in the previous sections, which reduces the number of transformer layers.

To ensure comparability between Depth-wise and Width-wise models, we scale the intermediate and attention layer dimensions in Width-wise Starbucks using a fixed division factor of six—matching the layer-step granularity used in Depth-wise configurations. Due to architectural constraints in BERT (e.g., the requirement for 12 attention heads), we further adjust the attention dimensions to ensure compatibility. Despite these adjustments, the parameter difference between corresponding Depth-wise and Width-wise models remains negligible—all under 0.2 million parameters ($<$0.05\%). Full configuration details are provided in Table~\ref{tab:width-wise-config} in the Appendix.

Table~\ref{tab:width-wise-results} presents a direct comparison between Depth-wise and Width-wise Starbucks variants. Before SMAE pre-training, both variants show comparable average effectiveness. However, after applying SMAE, Width-wise Starbucks consistently outperforms Depth-wise Starbucks, particularly for smaller model sizes.
We hypothesize that this improvement arises from the different types of contextualized representations induced by depth versus width reduction. Since these configurations alter the computation graph in distinct ways, their resulting embeddings are not functionally equivalent—especially in smaller sub-networks (e.g., Demi and Short). 

Motivated by this observation, we explore a \textbf{hybrid approach} that fuses embeddings from paired Depth-wise and Width-wise models of the same size. As shown in Table~\ref{tab:width-wise-results}, this approach provides the largest relative improvements for the smallest models, confirming the complementary nature of the two variants. As model size increases, the relative benefit of hybridization diminishes, suggesting that larger sub-networks already capture sufficiently rich representations. Overall, this hybrid approach consistently improves performance across all model sizes. Importantly, because Depth-wise and Width-wise Starbucks configurations operate independently, the hybridization can be performed in parallel with minimal additional inference cost. This makes the approach practical for real-world deployment scenarios where both effectiveness and efficiency are critical.

\section{Conclusion}

This paper highlights a key limitation of 2D Matryoshka Embedding models: despite their support for elastic embedding generation across layers and dimensions, their effectiveness remains well below that of separately trained models.

To address this, we propose \textbf{Starbucks}, a training framework that combines structured fine-tuning with masked autoencoding pre-training. Experiments on in-domain and out-of-domain tasks show that Starbucks outperforms 2D Matryoshka and matches—or even exceeds—the effectiveness of individually trained models, while remaining efficient and adaptable. Starbucks also yields a robust BERT-based backbone via SMAE, suitable for initializing downstream sub-models, with or without SRL. Ablation studies confirm the value of SMAE and demonstrate generalization to modern architectures like E5. Additionally, depth-wise and width-wise variants offer complementary representations, and a simple hybrid strategy further boosts performance with minimal latency overhead.

Overall, Starbucks bridges the effectiveness gap in elastic embedding models compared to separately trained small models, while retaining adaptability. It also opens promising directions for future work, such as automatically selecting optimal layer-dimension configurations to balance performance and efficiency in practical applications like embedding-as-a-service systems.
\newpage
\section{Limitations}

Our experiments focus on encoder-based dense retrievers, primarily using further pre-trained BERT and E5 checkpoints as embedding backbones. As a result, decoder-based models—such as RepLlama~\cite{repllama}, PromptReps~\cite{zhuang2024promptrepspromptinglargelanguage}, and related approaches—are not explored in this work. Investigating how our framework can be extended to or combined with decoder-based retrieval models remains a promising direction for future research.

The Starbucks configurations we explored include six representative settings (i.e., $(2, 32), (4, 64), \ldots$), chosen to balance variations in depth and width. While not exhaustive, our results show that increasing both the number of layers and embedding dimensions consistently improves performance. This suggests that further gains may be possible by identifying more optimal pairings of these parameters. Additionally, we did not evaluate configurations where the embedding dimension is fixed and only the depth varies—highlighting another avenue for exploring trade-offs and architectural flexibility.

Finally, while our study investigates Depth-wise and Width-wise Starbucks independently, a natural extension is to explore \textbf{3D Matryoshka models} that vary depth, width, and embedding dimension simultaneously. Such models could offer finer-grained control over efficiency-effectiveness trade-offs and enable dynamic adaptation across all axes of the architecture. This may provide greater deployment flexibility and enhance representational diversity by capturing complementary inductive biases. We leave the design and evaluation of such 3D configurations as exciting future work.




\bibliography{custom}

\appendix

\section{Appendix}\label{sec:appendix}

\subsection{Online Latency Comparison: Starbucks-sizes vs. Full Model}

\begin{figure}[htbp]
	\centering
	\includegraphics[width=\columnwidth]{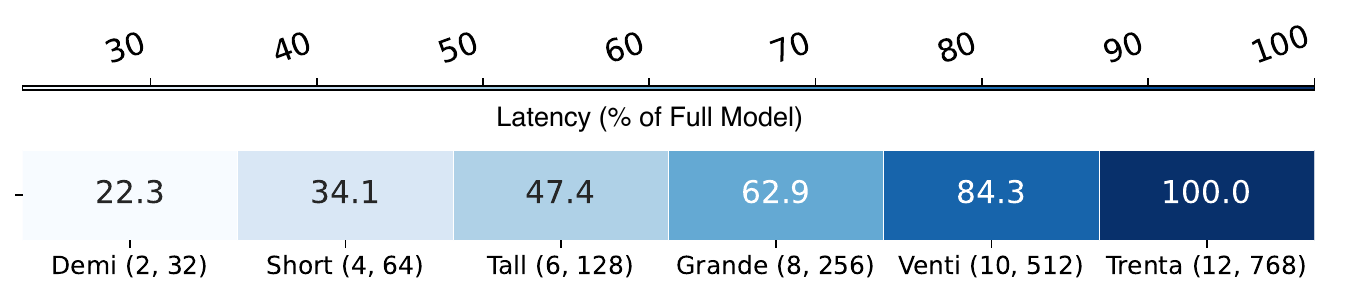}
	\caption{Latency comparison in percentage between Starbuck model sizes and a full BERT-based DPR model.}
	\label{fig:efficiency-heatmap}
\end{figure}
To quantify the efficiency benefits of Starbucks-sized sub-models, we measure their online latency relative to a full model. The comparison is based on the Dense Passage Retrieval (DPR) architecture using a BERT-base encoder, where each query and passage is independently encoded and similarity is computed via a dot product between their final embeddings. This setup follows prior work in dense retrieval.

As shown in Figure~\ref{fig:efficiency-heatmap}, we report the relative latency for each Starbucks size, averaged over 10{,}000 queries and 8.8 million passages—comparable in scale to the MS MARCO v1 corpus.

\subsection{Fine-tuning Hyperparameters}\label{appendix:parameters}
We report the fine-tuning hyperparameters used for the STS tasks and retrieval tasks in Tables~\ref{tab:ft_params_STS} and~\ref{tab:ft_params_Retrieval}, respectively.
%
%

\begin{table}[htbp]\centering
	\caption{Fine-tuning Hyperparameters for STS.}
	\label{tab:ft_params_STS}
			\resizebox{\columnwidth}{!}{
	\begin{tabular}{lc} 
		\toprule
		\bfseries Hyperparameter & \bfseries Assignment  \\
		\midrule
		learning Rate & 5e-5\\
		lr scheduler type & linear \\
		warmup ratio & 0.1 \\
		overall batch size & 128 \\
		optimizer & AdamW \\
		epochs & 1 \\
		fp16 & True \\
		kl\_temperature & 0.3 \\
		distance metric & cosine similarity \\
		Inner loss function & MultipleNegativesRankingLoss \\
		GPU & 1 x H100 80GB \\
		\bottomrule
	\end{tabular}
}
\end{table}

\begin{table}[htbp]\centering
\caption{Fine-tuning hyperparameters for passage retrieval.}
\label{tab:ft_params_Retrieval}
\begin{tabular}{lc} 
	\toprule
	\bfseries Hyperparameter & \bfseries Assignment  \\
	\midrule
	learning Rate & 1e-4 \\
	lr scheduler type & linear \\
	warmup ratio & 0.1 \\
	overall batch size & 128 \\
	optimizer & AdamW \\
	epochs & 3 \\
	fp16 & True \\
	kl\_temperature & 1.0 \\
	Inner loss function & InfoNCE \\
	distance metric & dot product \\
	\# hard negative & 7 \\
	in-batch negatives & True \\
	GPU & 1 x H100 80GB \\
	\bottomrule
\end{tabular}

\end{table}

\subsection{Full Results across Starbucks Sizes}\label{appendix:full_results}

\begin{table*}[h]
	\centering
	\scriptsize
	\caption{Complete results across all Starbucks sizes for in-domain datasets. Spearman's Correlation is used to evaluate the STS task. For retrieval task, MRR@10 is used to evaluate on MS MARCO dev set, while nDCG@10 is used to evaluate on DL19 and DL20. * denotes statistical significance differences (p<0.05) between the corresponding method and SMAE-SRL (Starbucks).}\label{table:full}
		\resizebox{\textwidth}{!}{
	\begin{tabular}{p{0.1pt}p{0.5pt}l|cccccccc|ccc}
		\toprule
		&&&\multicolumn{8}{c|}{\textbf{Semantic Text Similarity Tasks}} &\multicolumn{3}{c}{\textbf{Retrieval Tasks}} \\ \midrule
		&&\textbf{Method} & \textbf{STS-B} & \textbf{STS12} & \textbf{STS13} & \textbf{STS14} & \textbf{STS15} & \textbf{STS16} & \textbf{SICK-R} & \textbf{Average} & \textbf{MARCO dev} & \textbf{DL19} & \textbf{DL20} \\
		\midrule
		\multirow{5}{*}{\rotatebox{90}{\textbf{Demi}}}&\multirow{5}{*}{\rotatebox{90}{\textbf{(n=2, d=32)}}}
		& BERT              & 0.5919* & 0.4700* & 0.6052* & 0.5424* & 0.6530* & 0.6397* & 0.5996* & 0.5860 & 0.0001* &0.0000*& 0.0000*\\
		&& BERT-Separate     & 0.7134 & 0.6637 & 0.6770* & 0.6374* & 0.7451* & 0.7059* & 0.7210* & 0.6948 & 0.2092* &0.4599*&0.4826 \\
		&& BERT-2DMSE        & 0.7197 & 0.6204* & 0.6747* & 0.6278* & 0.7336* & 0.7058* & 0.7066* & 0.6841 &0.1589*&0.3459*&0.3828* \\
		&& BERT-SRL          & 0.7413 & 0.6789 & 0.7190 & 0.6684* & 0.7655* & 0.7309 & 0.7455 & 0.7214&0.2147*&0.5003&\textbf{0.5294} \\
		\cmidrule{3-14}
		&& SMAE-SRL (Starbucks) & \textbf{0.7455} & \textbf{0.6854} & \textbf{0.7368} & \textbf{0.6955} & \textbf{0.7972} & \textbf{0.7459} & \textbf{0.7473} & \textbf{0.7362} &\textbf{0.2282}&\textbf{0.5122}&0.5042\\
		\midrule
		
		\multirow{5}{*}{\rotatebox{90}{\textbf{Short}}}&\multirow{5}{*}{\rotatebox{90}{\textbf{(n=4, d=64)}}}
		& BERT             & 0.6021* & 0.4478* & 0.5881* & 0.5230* & 0.6599* & 0.6455* & 0.6083* &  0.5821 & 0.0000*&0.0000*&0.0000* \\
		&& BERT-Separate     & 0.7399* & 0.6882* & 0.7387* & 0.6800* & 0.7819* & 0.7419* & 0.7579* & 0.7326 & 0.2692* &0.5527&0.5487*\\
		&& BERT-2DMSE         & 0.7311* & 0.6364* & 0.6977* & 0.6551* & 0.7637* & 0.7312* & 0.7323* & 0.7068 & 0.2266*&0.4988*&0.5179* \\
		&& BERT-SRL           & 0.7621* & 0.7126 & 0.7400* & 0.6900* & 0.7924* & 0.7613 & 0.7678 & 0.7466&0.2722*&0.5653&0.5853 \\
		\cmidrule{3-14}
		&& SMAE-SRL (Starbucks) & \textbf{0.7933} & \textbf{0.7296} & \textbf{0.7824} & \textbf{0.7444} & \textbf{0.8308} & \textbf{0.7840} & \textbf{0.7700} & \textbf{0.7764}&\textbf{0.2950}&\textbf{0.5766}&\textbf{0.5920} \\
		\midrule
		
		\multirow{5}{*}{\rotatebox{90}{\textbf{Tall}}}&\multirow{5}{*}{\rotatebox{90}{\textbf{(n=6, d=128)}}}
		& BERT           & 0.6128* & 0.4689* & 0.6037* & 0.5464* & 0.6728* & 0.6624* & 0.6271* &  0.5991 & 0.0000*&0.0000*&0.0000* \\
		&& BERT-Separate     & 0.7853* & 0.7150 & 0.7794 & 0.7261* & 0.7959* & 0.7778 & 0.7802 & 0.7657 & 0.2935* &0.5939&0.6028 \\
		&& BERT-2DMSE         & 0.7501* & 0.6434* & 0.7061* & 0.6630* & 0.7739* & 0.7421* & 0.7522* &  0.7187 & 0.2532*&0.5464*&0.5503* \\
		&& BERT-SRL          & 0.7838* & 0.7158 & 0.7685* & 0.7146* & 0.8066* & 0.7762 & 0.7792 & 0.7635 &0.2983*&0.6102&0.6247\\
		\cmidrule{3-14}
		&& SMAE-SRL (Starbucks) & \textbf{0.8202} & \textbf{0.7374} & \textbf{0.7998} & \textbf{0.7654} & \textbf{0.8426} & \textbf{0.8029} & \textbf{0.7831} & \textbf{0.7931} &\textbf{0.3274}&\textbf{0.6346}&\textbf{0.6319}\\
		\midrule
		
		\multirow{5}{*}{\rotatebox{90}{\textbf{Grande}}}&\multirow{5}{*}{\rotatebox{90}{\textbf{(n=8, d=256)}}}
		& BERT              & 0.6433* & 0.5121* & 0.6370* & 0.5712* & 0.7062* & 0.6737* & 0.6504* & 0.6277 & 0.0009*&0.0000*&0.0000* \\
		&& BERT-Separate    & 0.8174 & 0.7304 & \textbf{0.8097} & 0.7588 & 0.8289* & 0.7998 & 0.7807 & 0.7894 & 0.3160*&0.6185&0.6259 \\
		&& BERT-2DMSE        & 0.7721* & 0.6635* & 0.7400* & 0.6756* & 0.7949* & 0.7570* & 0.7774 & 0.7401 &0.2770*&0.5604*&0.5856* \\
		&& BERT-SRL      &     0.8112 & 0.7310 & 0.7925 & 0.7397* & 0.8246* & 0.7959 & \textbf{0.7879} & 0.7833 &0.3159*&0.6342&\textbf{0.6344}\\
		\cmidrule{3-14}
		&& SMAE-SRL (Starbucks) & \textbf{0.8278} & \textbf{0.7382} & 0.8033 & \textbf{0.7736} & \textbf{0.8469} & \textbf{0.8092} & 0.7874 & \textbf{0.7981} &\textbf{0.3369}&\textbf{0.6525}&0.6292\\
		\midrule
		
		\multirow{5}{*}{\rotatebox{90}{\textbf{Venti}}}&\multirow{5}{*}{\rotatebox{90}{\textbf{(n=10, d=512)}}}
		& BERT              & 0.7331* & 0.6748* & 0.6983* & 0.6488* & 0.7876* & 0.7047* & 0.7133* & 0.7087 & 0.0000*&0.0100*&0.0047* \\
		&& BERT-Separate     & 0.8259 & 0.7379 & \textbf{0.8088} & 0.7691 & 0.8412 & 0.7931 & 0.7866 &0.7946 & 0.3279* &\textbf{0.6502}&0.6133\\
		&& BERT-2DMSE       & 0.7850* & 0.7097* & 0.7422* & 0.6927* & 0.8218* & 0.7626* & 0.7832 &  0.7568 & 0.3022* &0.5769*&0.6082\\
		&& BERT-SRL        & 0.8196 & 0.7346 & 0.8070 & 0.7635 & 0.8391 & 0.8003 & \textbf{0.7903} & 0.7935*&0.3249*&0.6405&\textbf{0.6499}\\
		\cmidrule{3-14}
		&& SMAE-SRL (Starbucks) & \textbf{0.8292} & \textbf{0.7405} & 0.7989 & \textbf{0.7772} & \textbf{0.8503} & \textbf{0.8138} & 0.7860 & \textbf{0.7994}&\textbf{0.3416}&0.6493&0.6310 \\
		\midrule
		
		\multirow{5}{*}{\rotatebox{90}{\textbf{Trenta}}}&\multirow{5}{*}{\rotatebox{90}{\textbf{(n=12, d=768)}}}
		& BERT    & \textbf{0.8411} & 0.7413 & \textbf{0.8319*} & \textbf{0.7902} & \textbf{0.8521} & \textbf{0.8151} & 0.7949* & \textbf{0.8095} &0.3341 &0.6404&0.6410 \\
		&& BERT-Separate    & \textbf{0.8411} & 0.7413 & \textbf{0.8319*} & \textbf{0.7902} & \textbf{0.8521} & \textbf{0.8151} & 0.7949* & \textbf{0.8095} &0.3341&0.6404&0.6410 \\
		&& BERT-2DMSE        & 0.8196 & 0.7414 & 0.8087 & 0.7616 & 0.8417 & 0.8017 & \textbf{0.7986*} & 0.7962 & 0.3185*&0.6021&0.6001*\\
		&& BERT-SRL          & 0.8258 & \textbf{0.7456} & 0.8185 & 0.7752 & 0.8474 & 0.8059 & 0.7889 & 0.8010&0.3266*&0.6550&\textbf{0.6518} \\
\cmidrule{3-14}
		&& SMAE-SRL (Starbucks) & 0.8274 & 0.7404 & 0.7998 & 0.7789 & 0.8497 & 0.8131 & 0.7844 & 0.7991&\textbf{0.3403}&\textbf{0.6558}&0.6358 \\
	\bottomrule
		
	\end{tabular}
}
\end{table*}

In Table~\ref{table:full}, we report the full results of all methods across various Starbucks sizes on in-domain datasets. We also indicate whether statistically significant differences ($p < 0.05$) were observed between each Starbucks configuration and the corresponding baselines. For retrieval tasks, we use a two-tailed paired Student’s $t$-test; for STS tasks, correlation coefficients are first transformed using Fisher’s $r$-to-$z$ transformation, followed by a $z$-test.

As the results show, Starbucks is the most effective method across most smaller configurations. For the full-size setting, separately trained models (BERT Separate) often perform best on STS tasks. Nevertheless, Starbucks remains the most effective Matryoshka-style training strategy overall in this setting.

Table~\ref{tab:beir_sizes} presents the results for out-of-domain datasets, evaluated using the BEIR benchmark. The results demonstrate that the findings from in-domain evaluation generalize well to out-of-domain settings.

\begin{table*}[htbp]
	\centering
	\scriptsize
	\caption{
	Complete results across all Starbucks sizes for out-of-domain datasets, evaluted using nDCG@10. * denotes statistical significance differences (p<0.05) between the corresponding method and SMAE-SRL (Starbucks).
}
	\label{tab:beir_sizes}
	\resizebox{\linewidth}{!}{
		\begin{tabular}{p{0.5pt}p{0.8pt}|l|cc|ccc|cc|c|cc|cc|c}
			\toprule
			&   & \textbf{Domain} & \multicolumn{2}{c|}{\textbf{General}} & \multicolumn{3}{c|}{\textbf{Scientific}} & \multicolumn{2}{c|}{\textbf{Debate}} & \multicolumn{1}{c|}{\textbf{Finance}} & \multicolumn{2}{c|}{\textbf{Biomedical}} & \multicolumn{2}{c|}{\textbf{Wiki}} & \textbf{Avg.} \\
			\midrule
			&   & \textbf{Dataset}& \textbf{Db.\-Ent.} & \textbf{Quora} & \textbf{Scidocs} & \textbf{Scifact} & \textbf{Cli.\-Fever} & \textbf{Arguana} & \textbf{Touche20} & \textbf{Fiqa} & \textbf{Trec\-Cvd} & \textbf{Nfcorpus} & \textbf{Fever} & \textbf{Hpqa} & \textbf{Avg.} \\
			
			\midrule
			\multirow{5}{*}{\rotatebox{90}{\textbf{Demi}}} & \multirow{5}{*}{\rotatebox{90}{\textbf{(n=2, d=32)}}} & BERT & 0.0011* & 0.0001* & 0.0013* & 0.0079* & 0.0000* & 0.0005* & 0.0000* & 0.0022* & 0.0000* & 0.0135* & 0.0000* & 0.0000* & 0.0022 \\
			&   & BERT-Separate & 0.1657 & 0.6900 & 0.0480* & 0.2692 & 0.0657* & 0.1517* & 0.1707* & 0.0832* & 0.3969* & 0.1607 & 0.3474* & 0.1620* & 0.2259 \\
			&   & BERT-2DMSE & 0.1314* & 0.5291* & 0.0414* & 0.2158* & 0.0582* & 0.1094* & 0.1601* & 0.0556* & 0.3262* & 0.1371* & 0.2111* & 0.1001* & 0.1730 \\
			&   & BERT-SRL & \textbf{0.1717} & 0.6865 & 0.0539 & 0.2577 & 0.0703* & 0.1471* & 0.1895* & 0.0806* & 0.3791* & 0.1622 & 0.3472* & 0.1698 & 0.2263 \\ \cmidrule{3-16}
			&   & SMAE-SRL (Starbucks) & 0.1680 & \textbf{0.6912} & \textbf{0.0602} & \textbf{0.2717} & \textbf{0.0977} & \textbf{0.1801} & \textbf{0.2336} & \textbf{0.1140} & \textbf{0.5051} & \textbf{0.1768} & \textbf{0.3630} & \textbf{0.1711} & \textbf{0.2527} \\
			\midrule
			\multirow{5}{*}{\rotatebox{90}{\textbf{Short}}} & \multirow{5}{*}{\rotatebox{90}{\textbf{(n=4, d=64)}}} & BERT & 0.0000* & 0.0001* & 0.0006* & 0.0000* & 0.0000* & 0.0012* & 0.0000* & 0.0000* & 0.0014* & 0.0100* & 0.0000* & 0.0001* & 0.0011 \\
			&   & BERT-Separate & 0.2205 & 0.7769* & 0.0741* & 0.3809 & 0.0912* & 0.2032* & 0.1890 & 0.1524* & 0.5089* & 0.2154 & 0.5084* & 0.3028* & 0.3020 \\
			&   & BERT-2DMSE & 0.1915* & 0.7157* & 0.0633* & 0.3122* & 0.0902* & 0.1851* & 0.1807* & 0.1164* & 0.4445* & 0.1811* & 0.4155* & 0.2241* & 0.2600 \\
			&   & BERT-SRL & 0.2227 & 0.7709* & 0.0758* & 0.3616 & 0.0990* & 0.2028* & 0.2074 & 0.1325* & 0.4838* & 0.2060* & 0.5009* & 0.2996* & 0.2969 \\ \cmidrule{3-16}
			&   & SMAE-SRL (Starbucks) & \textbf{0.2353} & \textbf{0.7829} & \textbf{0.0858} & \textbf{0.3857} & \textbf{0.1093} & \textbf{0.2273} & \textbf{0.2220} & \textbf{0.1859} & \textbf{0.6180} & \textbf{0.2270} & \textbf{0.5243} & \textbf{0.3159} & \textbf{0.3266} \\
			\midrule
			\multirow{5}{*}{\rotatebox{90}{\textbf{Tall}}} & \multirow{5}{*}{\rotatebox{90}{\textbf{(n=6, d=128)}}} & BERT & 0.0010* & 0.0821* & 0.0017* & 0.0087* & 0.0000* & 0.0048* & 0.0000* & 0.0004* & 0.0126* & 0.0142* & 0.0000* & 0.0000* & 0.0105 \\
			&   & BERT-Separate & 0.2498* & 0.8054* & 0.0929* & 0.4074 & 0.1122* & 0.2294* & 0.2306 & 0.1649* & 0.5527* & 0.2191* & 0.5786* & 0.3718* & 0.3346 \\
			&   & BERT-2DMSE & 0.2029* & 0.7587* & 0.0692* & 0.3492* & 0.1070* & 0.2102* & 0.2142 & 0.1432* & 0.4644* & 0.1918* & 0.4638* & 0.2669* & 0.2868 \\
			&   & BERT-SRL & 0.2443* & 0.7997* & 0.0899* & 0.4154 & 0.1214* & 0.2295* & 0.2210 & 0.1687* & 0.5619* & 0.2185* & 0.5495* & 0.3580* & 0.3315 \\ \cmidrule{3-16}
			&   & SMAE-SRL (Starbucks) & \textbf{0.2740} & \textbf{0.8150} & \textbf{0.1054} & \textbf{0.4333} & \textbf{0.1304} & \textbf{0.2537} & \textbf{0.2402} & \textbf{0.2202} & \textbf{0.6451} & \textbf{0.2609} & \textbf{0.5967} & \textbf{0.4044} & \textbf{0.3649} \\
			\midrule
			\multirow{5}{*}{\rotatebox{90}{\textbf{Grande}}} & \multirow{5}{*}{\rotatebox{90}{\textbf{(n=8, d=256)}}} & BERT & 0.0020* & 0.3888* & 0.0034* & 0.0494* & 0.0084* & 0.0461* & 0.0000* & 0.0030* & 0.0520* & 0.0419* & 0.0006* & 0.0015* & 0.0498 \\
			&   & BERT-Separate & 0.2820 & \textbf{0.8260*} & 0.1020* & \textbf{0.4688} & 0.1289* & 0.2416* & 0.2137 & 0.2077* & 0.5650* & 0.2396* & \textbf{0.6321*} & 0.4238* & 0.3609 \\
			&   & BERT-2DMSE & 0.2223* & 0.5341* & 0.0724* & 0.3605* & 0.1091* & 0.2237* & 0.2342 & 0.1558* & 0.5109* & 0.1993* & 0.5088* & 0.2954* & 0.2855 \\
			&   & BERT-SRL & 0.2729* & 0.8220* & 0.0992* & 0.4600 & 0.1329* & 0.2449* & 0.2202 & 0.1971* & 0.6103 & 0.2380* & 0.6056* & 0.4081* & 0.3593 \\ \cmidrule{3-16}
			&   & SMAE-SRL (Starbucks) & \textbf{0.2920} & 0.7808 & \textbf{0.1168} & 0.4538 & \textbf{0.1430} & \textbf{0.2651} & \textbf{0.2451} & \textbf{0.2379} & \textbf{0.6672} & \textbf{0.2719} & 0.6243 & \textbf{0.4380} & \textbf{0.3780} \\
			\midrule
			\multirow{5}{*}{\rotatebox{90}{\textbf{Venti}}} & \multirow{5}{*}{\rotatebox{90}{\textbf{(n=10, d=512)}}} & BERT & 0.0005* & 0.1195* & 0.0040* & 0.0250* & 0.0000* & 0.0085* & 0.0000* & 0.0000* & 0.0138* & 0.0247* & 0.0000* & 0.0000* & 0.0163 \\
			&   & BERT-Separate & \textbf{0.3109} & 0.7687* & 0.1087* & 0.4665 & 0.1376* & 0.2646 & 0.2299 & 0.2277* & 0.5898* & 0.2543* & \textbf{0.6666*} & \textbf{0.4652*} & \textbf{0.3742} \\
			&   & BERT-2DMSE & 0.2442* & 0.1566* & 0.0807* & 0.3560* & 0.1305* & 0.2301* & 0.2466 & 0.1791* & 0.5437* & 0.2030* & 0.5687* & 0.3267* & 0.2722 \\
			&   & BERT-SRL & 0.2856 & \textbf{0.7751*} & 0.1070* & \textbf{0.4727} & 0.1485 & 0.2553* & 0.2486 & 0.2206* & 0.6103* & 0.2460* & 0.6430* & 0.4439 & 0.3714 \\ \cmidrule{3-16}
			&   & SMAE-SRL (Starbucks) & 0.2968 & 0.0178 & \textbf{0.1175} & 0.4629 & \textbf{0.1538} & \textbf{0.2673} & \textbf{0.2545} & \textbf{0.2486} & \textbf{0.6645} & \textbf{0.2733} & 0.6359 & 0.4466 & 0.3200 \\
			\midrule
			\multirow{5}{*}{\rotatebox{90}{\textbf{Trenta}}} & \multirow{5}{*}{\rotatebox{90}{\textbf{(n=12, d=768)}}} & BERT & \textbf{0.3018} & 0.8224* & 0.1109* & 0.4597 & 0.1604 & 0.2707 & 0.2337* & 0.2200* & 0.6041 & 0.2585* & \textbf{0.6601*} & \textbf{0.4548*} & 0.3798 \\
			&   & BERT-Separate & \textbf{0.3018} & 0.8224* & 0.1109* & 0.4597 & 0.1604 & 0.2707 & 0.2337* & 0.2200* & 0.6041 & 0.2585* & \textbf{0.6601*} & \textbf{0.4548*} & 0.3798 \\
			&   & BERT-2DMSE & 0.2552* & 0.8152* & 0.0878* & 0.3863* & 0.1391* & 0.2363* & 0.1936 & 0.1851* & 0.5673 & 0.2105* & 0.5848* & 0.3532* & 0.3345 \\
			&   & BERT-SRL & 0.2925 & 0.3419* & 0.1072* & \textbf{0.4826} & 0.1569 & 0.2533* & \textbf{0.2628*} & 0.2217* & \textbf{0.6294} & 0.2508* & 0.6470* & 0.4466 & 0.3411 \\ \cmidrule{3-16}
			&   & SMAE-SRL (Starbucks) & 0.2997 & \textbf{0.8477} & \textbf{0.1206} & 0.4715 & \textbf{0.1618} & \textbf{0.2734} & 0.2022 & \textbf{0.2508} & 0.6140 & \textbf{0.2738} & 0.6275 & 0.4486 & \textbf{0.3826} \\
			\bottomrule
		\end{tabular}
	}
\end{table*}

%
%
%
%
%
%
%
%

\subsection{Separate Trained Models with SMAE}

\begin{figure}[htbp]
	\includegraphics[width=\linewidth]{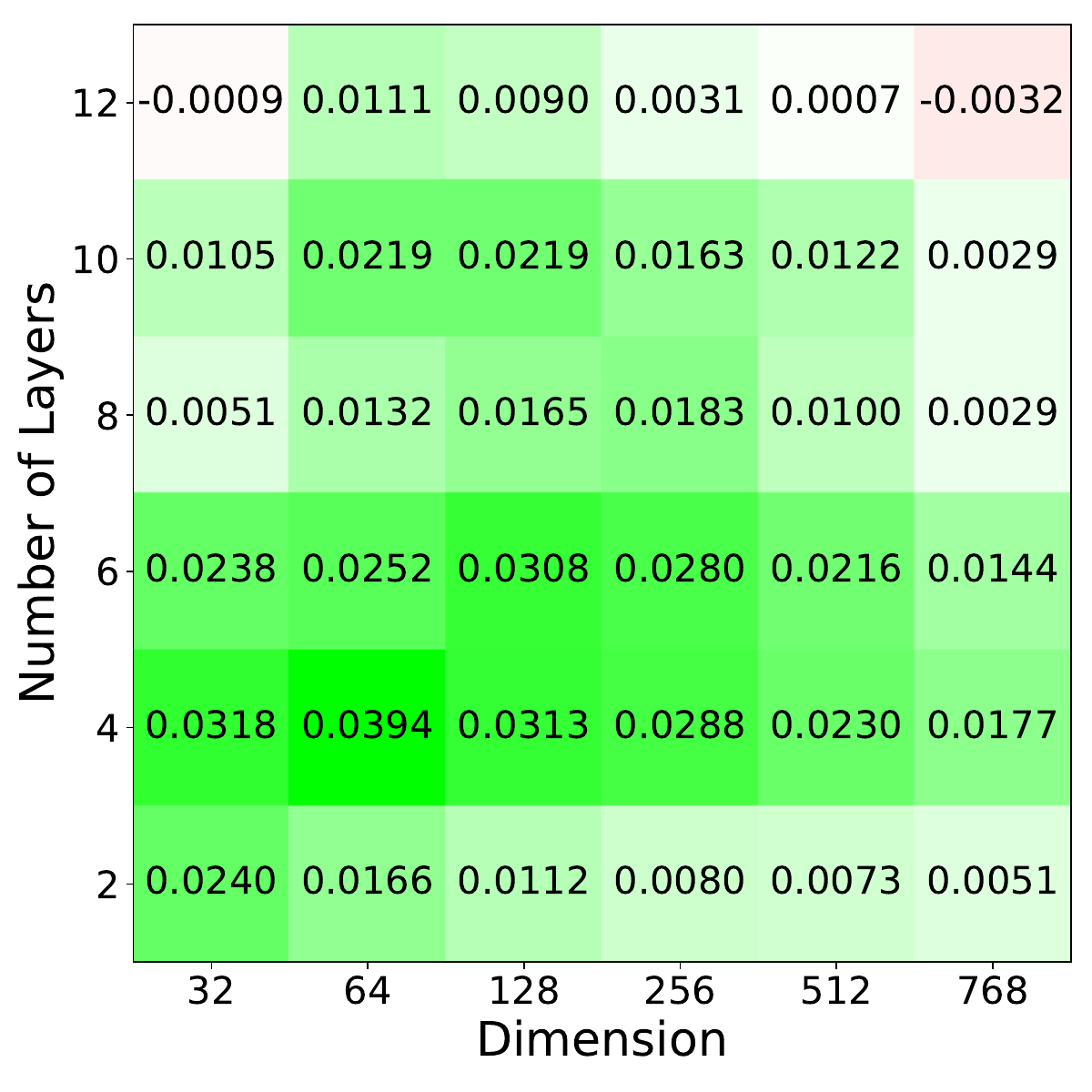}
	\caption{Average Spearman's Correlation deltas between separately trained models with SMAE and BERT (i.e. SMAE-Separate - BERT-Separate) on seven STS datasets for different layer and dimensionality sizes. Negative values (red) indicate the separate trained model was better.}
	\label{fig:starbucks_vs_baseline}
\end{figure}

Figure~\ref{fig:starbucks_vs_baseline} shows the comparison of separately trained models across different sizes using either SMAE or BERT on STS tasks. Although SMAE pre-training specifically targets settings in the Starbucks diagonal, it also benefits other combinations of dimensions and layers.
 
\subsection{Width-wise Starbuck Sizes}

In addition to reducing model depth, Starbucks supports width reduction by shrinking the size of intermediate and attention layers. This enables a more efficient encoder while preserving the number of layers. Table~\ref{tab:width-wise-config} shows the specific configurations (i.e specific dimensions) we use for Width-wise Starbucks across different Starbuck sizes.

\begin{table}[htbp]\centering
	\caption{Configuration of Width-wise Starbucks variants with reduced intermediate and attention layer sizes.}
	\label{tab:width-wise-config}
	\resizebox{\linewidth}{!}{
	\begin{tabular}{lcc}
		\toprule
		\textbf{Model Size} & \textbf{Intermediate Dim} & \textbf{Attention Dim} \\
		\midrule
		Demi   & 512   & 132 \\
		Short  & 1024  & 252 \\
		Tall   & 1536  & 384 \\
		Grande & 2048  & 516 \\
		Venti  & 2560  & 636 \\
		Trenta & 3072  & 768 \\
		\bottomrule
	\end{tabular}
}
\end{table}

\end{document}